\documentclass{tut2016} % put ./tut2016 if using only locally
\usepackage[colorlinks=true,breaklinks=true,bookmarks=false,urlcolor=blue,citecolor=blue,linkcolor=blue,bookmarksopen=false,draft=false]{hyperref}
\hypersetup{hidelinks}
\usepackage[symbol]{footmisc}
\usepackage{amsmath}
\usepackage{amssymb}
\usepackage[bbgreekl]{mathbbol}
\DeclareSymbolFontAlphabet{\mathbb}{AMSb}
\usepackage{titlesec}
\usepackage{fancyhdr}
\usepackage{color, colortbl}
\usepackage{accents}
\usepackage{floatrow}
\usepackage{caption}
\usepackage{bm}
\usepackage{tikz}
\usetikzlibrary{calc}
\usetikzlibrary{matrix, positioning}
\usepackage[square,sort,comma,numbers]{natbib}
\bibpunct[, ]{(}{)}{,}{a}{}{,}%

\makeatletter

\newtheorem{subassumption}{Assumption\normalfont}[assumption]

\newenvironment{assumption+}[1]
 {\renewcommand{\thesubassumption}{#1}\subassumption}
 {\endsubassumption}

\newcommand{\mathbbm}[1]{\text{\usefont{U}{bbm}{m}{n}#1}} % from mathbbm.sty
\newcommand{\bI}{\mathbbm{1}}
\newcommand{\independent}{\perp \!\!\! \perp}
\newcommand{\Var}{\mathrm{Var}}
\newcommand{\Bias}{\mathrm{Bias}}
\newcommand{\Cov}{\mathrm{Cov}}
\newcommand{\cF}{\mathcal{F}}
\newcommand{\cP}{\mathcal{P}}
\newcommand{\bE}{\mathrm{E}}
\newcommand{\bR}{\mathbb{R}}
\newcommand{\cY}{\mathcal{Y}}
\newcommand{\cG}{\mathcal{G}}
\newcommand{\cW}{\mathcal{W}}
\newcommand{\cX}{\mathcal{X}}
\newcommand{\cN}{\mathcal{N}}
\newcommand{\bb}{\mathbbm}

\newcommand\blfootnote[1]{%
  \begingroup
  \renewcommand\thefootnote{}\footnote{#1}%
  \addtocounter{footnote}{-1}%
  \endgroup
}

\begin{document}
%%%%%%%%%%%%%%%%

\CHAPTERNO{}% Enter chapter number here, e.g. "Chapter 1"
\TITLE{Experimental Design For Causal Inference Through An Optimization Lens}    % Enter chapter title here
%\SUBTITLE{}% Enter substitle (only if enecessary) and outcomment

% Enter author(s) here:
\AUBLOCK{% 
  \AUTHOR{Jinglong Zhao}
  \AFF{Boston University, Questrom School of Business, % \EMAIL is a part of \AFF(iliation) 
       \EMAIL{jinglong@bu.edu}
  }
  % ... up to last author
} % This ends \AUBLOCK

\CHAPTERHEAD{Experimental Design for Causal Inference}% Shortened running head "Author(s): Short Title"

\ABSTRACT{%
The study of experimental design offers tremendous benefits for answering causal questions across a wide range of applications, including agricultural experiments, clinical trials, industrial experiments, social experiments, and digital experiments.
Although valuable in such applications, the costs of experiments often drive experimenters to seek more efficient designs.
Recently, experimenters have started to examine such efficiency questions from an optimization perspective, as experimental design problems are fundamentally decision-making problems. 
This perspective offers a lot of flexibility in leveraging various existing optimization tools to study experimental design problems. 
This tutorial thus aims to examine the foundations of experimental design problems in the context of causal inference as viewed through an optimization lens.
% put abstract here (required)
} % This ends the abstract

\KEYWORDS{%
Experimental design, causal inference, optimization
% enter keywords here (required)
} % This ends the keywords

% \HISTORY{\href{https://papers.ssrn.com/sol3/papers.cfm?abstract_id=4780792}{First draft: April 1, 2024.} This version: \today}

%%%%%%%%%%%
% \maketitle  % This will make the chapter opening ("titlepage")
%%%%%%%%%%%

\begin{center}
{\Large\bf Experimental Design For Causal Inference} 
\vskip 0.5em
{\Large\bf Through An Optimization Lens}
\vskip 1.5em
Jinglong Zhao \blfootnote{Jinglong Zhao, Questrom School of Business, Boston University, \url{jinglong@bu.edu}. 
The author thanks Lihua Lei for sharing valuable perspectives that built the foundation of this manuscript and Alberto Abadie, David Alderson, Volodymyr Babich, Yuchen Hu, Stefanus Jasin, Ramesh Johari, Shuangning Li, Tu Ni, Chao Qin, Christopher Ryan, Nian Si, Johan Ugander, Ruoxuan Xiong, Ruohan Zhan, Bo Zhang, course participants at the National University of Singapore, and five anonymous referees for suggestions and feedbacks that significantly improved this manuscript. 
This manuscript builds on the author's lecture notes taught at the National University of Singapore.}\\
Boston University\\
\vskip 1em%
{\today} \par
\vskip 2em%
\end{center}\par

\begin{center} \normalsize\bf\text{Abstract} \end{center}

\begin{quote}\normalsize
The study of experimental design offers tremendous benefits for answering causal questions across a wide range of applications, including agricultural experiments, clinical trials, industrial experiments, social experiments, and digital experiments.
Although valuable in such applications, the costs of experiments often drive experimenters to seek more efficient designs.
Recently, experimenters have started to examine such efficiency questions from an optimization perspective, as experimental design problems are fundamentally decision-making problems. 
This perspective offers a lot of flexibility in leveraging various existing optimization tools to study experimental design problems. 
This manuscript thus aims to examine the foundations of experimental design problems in the context of causal inference as viewed through an optimization lens.
\end{quote}

\vskip 2em%

\section{Introduction}
\label{sec:Intro}

The study of experimental design offers tremendous benefits for answering causal questions across a wide range of applications, including agricultural experiments, clinical trials, industrial experiments, social experiments, and digital experiments.
Experimental design is probably one of the cleanest ways to answer causal questions, that is, to understand the causation behind some phenomenon.
In such an experiment, the experimenter usually compares the standard offering of some existing policy (the ``control'') to a new version of the policy (the ``treatment'') by splitting the experimental units into the control and treatment groups. 
By comparing the outcomes from these two groups, the experimenter discovers the ``treatment effect,'' that is, the degree to which the newer version is better or worse than the standard version.
See Examples~\ref{exa:RCTCows}~--~\ref{exa:RCTUber} below for an illustration of experimental design terminology in the context of causal inference.

\begin{example}[\citet{cochran1941double}. Agricultural Experiments]
\label{exa:RCTCows}
During the winter of 1939 - 1940, the Iowa Agricultural Experiment Station (AES) conducted feeding experiments on their 18 Holstein cows to compare three feeding plans on milk yield.
For simplicity, we drop one feeding plan from \citet{cochran1941double} and only compare two feeding plans in this example.
We refer to the plan of feeding roughage (alfalfa hay and corn silage) as the control and the plan of feeding grain plus roughage as the treatment.

We present a simplified version of the experiment that Iowa AES conducted.
On each day, Iowa AES randomly allocated these 18 Holstein cows into either the treatment or the control group.
We refer to each cow as an experimental unit.
At the end of each day, Iowa AES measured the average milk yields of the two groups of cows.
We refer to the difference between the two milk yields as the treatment effect.
\end{example}

\begin{example}[\citet{landovitz2021cabotegravir}. Clinical Trials]
\label{exa:RCTHIV}
In May 2020, the HIV Prevention Trials Network (HPTN) concluded a randomized clinical trial on cabotegravir (a medicine) for human immunodeficiency virus (HIV) prevention in seven African countries.
We refer to the long-acting cabotegravir as the treatment and the FDA-approved standard therapy (daily oral tenofovir disoproxil fumarate-emtricitabine) as the control.
Note that such a control is usually referred to as an ``active control,'' that is, a control that is not a placebo.
See \citet{he2024generalizing} for a detailed introduction of active control trials.

This landmark clinical trial was conducted on 4566 participants (ignoring dropouts for a simpler exposition in this example) over a duration of 153 weeks.
Participants had regular site visits and had their HIV status tested. 
We refer to each participant as an experimental unit.
After the trial, HPTN measured the time-to-HIV diagnosis outcomes on the two groups of participants, which can be thought of as two rates that indicate the effect of prevention.
We refer to the difference in rates between the two groups as the treatment effect.
\end{example}

\begin{example}[\citet{maruthi1999improving}. Industrial Experiments]
\label{exa:RCTCircuitBoard}
In 1999, Indal Electronics Ltd. considered using randomized experiments to reduce defects and enhance yields in the manufacturing of high, dense, inner-layer circuits for printed circuit board (PCB).
The experiment evaluated various parameters of the manufacturing process, including surface preparation, lamination conditions, exposure steps, and etching parameters.
For simplicity, we consider two configurations and refer to the new configuration of these process parameters as the treatment and the existing configuration as the control.

Conducted over a six-month production period, this industrial experiment is projected to generate annual savings of approximately \$50,000 for the company. 
In this experiment, we refer to each PCB as an experimental unit.
After the experiment, defects were assessed by measuring occurrences of shorts (i.e., unintended connections between conductors) and opens (i.e., unintended disconnections) in the PCBs. 
We refer to the difference in the number of defects as the treatment effect.
\end{example}

\begin{example}[\citet{mosteller1995tennessee}. Social Experiments]
\label{exa:RCTSTAR}
Between 1985 and 1989, the educational system of Tennessee conducted an experiment called Project STAR (Student-Teacher Achievement Ratio) to assess the effectiveness of small-sized classes compared to regular-sized classes on students' cognitive achievements.
For simplicity, we drop one type of classes and only compare two types in this example.  
We refer to the small-sized classes with 13-17 students as the treatment and the regular-sized classes with 22-25 students as the control.

This landmark educational experiment involved approximately 6500 students at 80 schools over a duration of four years (in kindergarten and in the first, second, and third grades).
We refer to each student as an experimental unit.
After the experiment, the educational system of Tennessee observed the standardized and curriculum-based test scores.
We refer to the difference in test scores as the treatment effect.
\end{example}

\begin{example}[\citet{farronato2018innovation}. Digital Experiments]
\label{exa:RCTUber}
In November 2017, Uber’s data scientists launched pilot experiments in Boston and San Francisco to test a number of surge pricing algorithms to their new product called Uber Express POOL.
In an experiment, we refer to the new surge pricing algorithm as the treatment and the status-quo flat rate pricing algorithm as the control.

Uber runs different types of experiments; among those most commonly used are user-level experiments, switchback experiments, and synthetic control experiments.
\begin{enumerate}
\itemsep0em
\item In a user-level experiment, Uber randomly allocates riders into either the treatment or the control group.
We refer to each rider as an experimental unit.
Uber typically includes millions of units in a user-level experiment.
After the experiment, Uber measures the ride-making frequencies of the two groups of riders. 
We refer to the difference between the two frequencies as the treatment effect.
\item In a switchback experiment, Uber focuses on the treatment effect at a given city such as Boston.
Rather than randomly splitting the user population, a switchback experiment randomly switches between treatment and control based on time.
Uber alternatively exposes all riders in Boston to treatment or control for a period of approximately 160 minutes. 
We refer to each 160-minute period as an experimental unit.
For a typical experiment of two to four weeks, hundreds of units are included in a switchback experiment.
\item In a synthetic control experiment, Uber compares the treatment effect across multiple cities.
Uber selects one city (or a few cities) and exposes all riders in this city to treatment; in the many more unselected cities, Uber exposes all riders to control.
Uber carefully selects the treatment city that is representative of the control cities, so that the treatment city can serve as a synthetic control city to compare with the actual control cities.
We refer to each city as an experimental unit.
At Uber, typically less than a hundred units, with only one or very few treatment units, are included in a synthetic control experiment.
\end{enumerate}
\end{example}

Other than the above examples, experimental design has been recognized as the gold standard in the new product development processes at technology firms \citep{koning2022experimentation}.
Practitioners from a variety of firms have reported their intensive usage of experiments in their product iterations \citep{chamandy2016experimentation, chen2024new, farias2023correcting, farronato2018innovation, gupta2019top, huang2023estimating, tang2020control, ye2023deep, zhu2024seller}, including search engines (e.g., Bing, Google, Yandex), online retailers (e.g., Amazon, eBay, Etsy), media services (e.g., Netflix), short-form video hosting services (e.g., Douyin, TikTok), social networking services (e.g., Facebook, LinkedIn, Twitter, WeChat), on-demand service platforms (e.g., DoorDash, Lyft, Uber), and travel services (e.g., Airbnb, Booking.com).
These firms conduct thousands of new experiments per week to innovate new products and accelerate their product iterations \citep{kohavi2013online, tang2010overlapping}.

Although experiments have proven beneficial across a wide range of applications, practical constraints such as market size, time, resources, and experimental risks may limit the ``sample size'' of an experiment --- that is, the total number of experimental units to allocate to the treatment and control groups.
Conducting experiments with excessively large sample sizes can be financially and logistically challenging.
An emerging question for experimenters, despite their millions and even billions of experimental units, is how to use the experimental units efficiently to draw accurate conclusions. 

In recent years, researchers have started to examine such efficiency questions in modern applications through an optimization lens. 
Experimental design problems are fundamentally decision-making problems \citep{bickel2015mathematical, casella2002statistical, neyman1933ix}. 
Given the objectives of the experimenter, the information and uncertainty faced by the experimenter, and the constraints that must be followed, the experimental design problems can be formulated as well-defined optimization problems. 
Such a perspective offers a lot of flexibility in using various existing optimization tools to study experimental design problems. 
This manuscript thus aims to examine the foundations of experimental design problems in the context of causal inference as viewed through an optimization lens.

\subsubsection*{Structure and scope.}

This manuscript is structured as follows. 
Section~\ref{sec:ExpDesign} introduces the key elements of experimental design for causal inference and outlines three major optimization frameworks derived from decision theory \citep{wald1949statistical, savage1951theory, lindley1956measure, wu1981robustness}.
We refer to these three frameworks as the robust optimization framework, the stochastic optimization framework, and the deterministic optimization framework.
Section~\ref{sec:Uncertainty1} illustrates the robust optimization framework in greater details.
Section~\ref{sec:Uncertainty2} introduces additional key elements related to a notion called ``covariates,'' and uses covariates to illustrate the stochastic optimization framework in greater details.
Section~\ref{sec:Uncertainty3} uses covariates to illustrate the deterministic optimization framework in greater details.
Section~\ref{sec:Survey} revisits three basic assumptions from Section~\ref{sec:ExpDesign}; the violation of each assumption leads to many active research directions in modern experimental design for causal inference literature.
Section~\ref{sec:Survey} picks one direction from the violation of each assumption and surveys recent developments in these three directions.
We conclude in Section~\ref{sec:Conclusion} with recommendations on how to choose the appropriate framework.
% The formal proofs to all of the lemmas and theorems are available from \citet{zhao2024experimental}.
This manuscript is self-contained, including proofs to all the lemmas and theorems, which can be found in the appendix.

Experimental design is a broad area; as such, this manuscript only covers a narrower scope on experimental design for causal inference.
One important omission is the rich literature on ``optimal experimental design,'' or ``optimal design'' for short, that draws from the literature of linear algebra, optimization, and statistics.
We only discuss the optimal experimental design literature when it overlaps with experimental design for causal inference in Section~\ref{sec:Uncertainty3}.
We refer to \citet{atkinson2007optimum, fedorov2013theory, pukelsheim2006optimal, silvey2013optimal} for textbooks and \citet{atkinson1975optimal, card1993minimum, titterington1975optimal} for papers on optimal experimental design.
A related omission is the literature on ``optimal Bayesian design.'' 
We only discuss the optimal Bayesian design literature when it overlaps with experimental design for causal inference in Section~\ref{sec:Uncertainty2}.
We refer to \citet{chaloner1982optimal, chaloner1995bayesian, kasy2016experimenters, letham2019constrained, lindley1972bayesian} for foundational papers and surveys on optimal Bayesian design.

Causal inference is also a broad area.
Here, two major approaches are used to answer causal inference questions: the ``potential outcomes'' approach and the ``structural causal model'' approach, and each provide complementary insights.
This manuscript only adopts the potential outcomes approach.
One important omission is the structural causal model approach.
For more details about the two approaches, we refer to \citet{ding2023first, hernan2010causal, imbens2015causal, rosenbaum2010design, wager2020stats} for textbooks on the potential outcomes approach,
and refer to \citet{pearl2000models, pearl2016causal, peters2017elements, spirtes2001causation} for textbooks on the structural causal model approach.

The mathematical foundation of this manuscript draws from optimization and statistics.
For a bigger picture of these two areas, we refer to \citet{ben2009robust, bertsekas1997nonlinear, bertsimas1997introduction, birge2011introduction, boyd2004convex, nocedal1999numerical, schrijver2003combinatorial} for textbooks on optimization,
and \citet{berger2013statistical, bickel2015mathematical, buhlmann2011statistics, casella2002statistical, chen2022elements, rigollet2019high, wainwright2019high, wooldridge2010econometric} for textbooks on statistics.

Throughout this manuscript, we will use capital letters (e.g., $W$) for random variables and lowercase letters (e.g., $w$) for deterministic quantities.
We will use regular font (e.g., $w$) for scalars, bold font (e.g., $\bm{w}$) for vectors, and blackboard font (e.g., $\bb{x}$, $\bbbeta$) for matrices.
Our notation for matrices is uncommon, yet it is used to distinguish between random variables and deterministic quantities.
Unless otherwise stated, we follow the convention that all vectors are column vectors.
For any vector $\bm{w}$ or matrix $\bb{x}$, we use $\bm{w}^\top$ and $\bb{x}^\top$ to stand for their transposes.

\section{Experimental Design for Causal Inference}
\label{sec:ExpDesign}

We start with one single experimental unit.
Let there be two versions of treatments, the active treatment (referred to as ``treatment'') and the controlled treatment (referred to as ``control''), which we denote using $1$ and $0$, respectively.
The random treatment assignment $W$ of this unit takes values from $\{0,1\}$.
Following convention, let $W$ stand for a random treatment assignment and $w$ stand for one realization.

One popular framework of causal inference is the potential outcomes framework, which was first proposed in \citet{neyman1923application} and subsequently studied by \citet{holland1986statistics, rubin1974estimating}. 
See \citet{athey2017econometrics, ding2023first, imbens2015causal} for more bibliographical notes.
The potential outcomes framework states that, for this single unit, there exists a pair of two random variables called ``potential outcomes,'' which each corresponds to a version of treatment.
Let $Y(1)$ and $Y(0)$ be the potential outcomes under the treatment assignment and under the control assignment, respectively.
Let $(Y(1), Y(0)) \sim \cF$ denote that the potential outcomes come from a joint distribution $\cF$.

Although there is a pair of two potential outcomes, we can only observe (by drawing a sample from) one of the two potential outcomes.
The observed outcome $Y$, sometimes also written as $Y^\mathsf{obs}$, is connected to the potential outcomes by
\begin{align*}
Y = Y(W) = \left\{ 
\begin{aligned}
Y(1), & \ \text{if} \ W = 1, \\
Y(0), & \ \text{if} \ W = 0.
\end{aligned}
\right.
\end{align*}
Note that, whenever we write $Y = Y(W)$ for two random variables, we mean that these two random variables are equal almost surely. 
The observed outcome $Y$ has two sources of randomness.
The first source of randomness comes from the potential outcomes $(Y(1), Y(0))$ as they are sampled from a joint distribution $\cF$;
the second source of randomness comes from the random treatment assignment $W$.

Under the potential outcomes framework, any comparison of potential outcomes has a causal interpretation.
One popular choice of the causal effect, or causal estimand, of interest is the average difference between the potential outcomes under treatment and control,
\begin{align*}
\bE_{\cF}\big[Y(1) - Y(0)\big].
\end{align*}
Because only one of the two potential outcomes can be observed and the other is always missing, the difference $Y(1) - Y(0)$ can never be directly observed.
This is also referred to as the fundamental challenge of causal inference by \citet{holland1988causal}.

Because of this fundamental challenge, the literature usually makes additional assumptions and relies on the availability of multiple units in making causal inference.
Now we generalize the above discussion to multiple units.
Let there be a total of $n$ units. 
We refer to $n$ as the ``sample size'' of an experiment.
The experimental units are indexed by $j \in [n] := \{1,2,...,n\}$.
For each unit $j \in [n]$, let the random treatment assignment be $W_j$, which takes values from $\{0,1\}$.
We collect all treatment assignments in a vector form as $\bm{W}$ that takes values from $\{0,1\}^n$.
Following convention, let $\bm{W}$ stand for a vector of random treatment assignments and $\bm{w}$ stand for one realization.

In the most general sense, for each unit $j \in [n]$, there are $2^n$ potential outcomes denoted as $Y_j(\bm{w})$ for all $\bm{w} \in \{0,1\}^n$.
We next introduce the non-interference assumption to simplify these many potential outcomes.

\begin{assumption}[Non-interference]
\label{asp:NonInterference}
For any two vectors of treatment assignments $\bm{w}, \bm{w}' \in \{0,1\}^n$, and for any unit $j \in [n]$, if $w_j = w'_j$, then 
\begin{align*}
Y_j(\bm{w}) = Y_j(\bm{w}').
\end{align*}
\end{assumption}

The non-interference assumption states that one unit's treatment assignment does not affect the outcomes of any other unit.
This assumption dates back to \citet{cox1958planning}, and serves as a critical component to the stable unit treatment value assumption (SUTVA), which was formally introduced in \citet{rubin1980discussion}.
% We refer to \citet{ding2023first} Section 2.2 and \citet{imbens2015causal} Section 1.6 for the full definition of SUTVA.
We refer to \citealp[Section 2.2]{ding2023first} and \citealp[Section 1.6]{imbens2015causal} for the full definition of SUTVA.
Assumption~\ref{asp:NonInterference} holds in many applications such as agricultural experiments, laboratory experiments, and clinical trials, when experiments are perfectly controlled.
Assumption~\ref{asp:NonInterference} may not always hold in other applications such as social experiments and marketplace experiments. 
Violation of Assumption~\ref{asp:NonInterference} leads to an active research direction on interference.
See Section~\ref{sec:Survey} for further discussions.

We make Assumption~\ref{asp:NonInterference} (i.e., the non-interference assumption) throughout this manuscript.
Under Assumption~\ref{asp:NonInterference}, for each unit $j \in [n]$, the observed outcome $Y_j$ is connected to a pair of two potential outcomes $(Y_j(1), Y_j(0))$ by
\begin{align*}
Y_j = Y_j(W_j) = \left\{ 
\begin{aligned}
Y_j(1), & \ \text{if} \ W_j = 1, \\
Y_j(0), & \ \text{if} \ W_j = 0.
\end{aligned}
\right.
\end{align*}
We collect all $2n$ potential outcomes as $(\bm{Y}(1), \bm{Y}(0))$, which takes values from $\cY \subseteq \bR^{2n}$.
\citet{ding2023first} refers to these $2n$ potential outcomes $(\bm{Y}(1), \bm{Y}(0))$ as the ``science table,'' a term adapted from \citet{rubin2005causal}.

Although Assumption~\ref{asp:NonInterference} greatly simplifies the potential outcomes, we would usually rely on a further assumption on the underlying data-generating process.

\begin{assumption}[Homogeneity]
\label{asp:iid}
For each unit $j \in [n]$, the pair of potential outcomes $(Y_j(1), Y_j(0))$ associated with unit $j$ is a pair of independent and identically distributed (i.i.d.) random variables that come from the same super-population, that is,
\begin{align*}
(Y_j(1), Y_j(0)) \sim \cF.
\end{align*}
\end{assumption}

Assumption~\ref{asp:iid} states that the units are independent and identically distributed. 
So the observed outcome of one unit is comparable to the observed outcome of any other unit (in the same treatment group).
Collecting the observed outcomes from all the units enables us to understand the underlying population.
Violation of Assumption~\ref{asp:iid} leads to many active research directions in the literature; one of these is treatment heterogeneity.
See Section~\ref{sec:Survey} for further discussions.

Under Assumption~\ref{asp:iid} and letting $(Y(1), Y(0)) \sim \cF$, we define the average causal effect of the population as
\begin{align}
\tau = \bE_{\cF} \big[Y(1) - Y(0)\big]. \label{eqn:tau:population}
\end{align}
Experimental designs that involve the above causal effect $\tau$, together with Assumption~\ref{asp:iid} or some other assumption on the underlying data-generating process, are sometimes referred to as taking a ``sampling-based'' perspective, so named because the potential outcomes are obtained by sampling from a super-population.
Under the sampling-based perspective, we essentially wish to understand the underlying population.

On the other hand, we could also focus on a finite sample and conduct the entire analysis by conditioning on the realized values of the $2n$ potential outcomes; that is, by conditioning on $(\bm{Y}(1), \bm{Y}(0)) = (\bm{y}(1), \bm{y}(0))$. 
% It is just that we can only observe $n$ observed outcomes out of the $2n$ potential outcomes.
Conditioning on the realized values of the $2n$ potential outcomes, we define the average causal effect of a finite sample as
\begin{align}
\tau_{\bm{y}(1), \bm{y}(0)} = \frac{1}{n} \sum_{j=1}^n \big(y_j(1) - y_j(0)\big). \label{eqn:tau:sample}
\end{align}
We make the distinction that whenever we write $\tau_{\bm{Y}(1), \bm{Y}(0)}$, this quantity is random in nature, as the potential outcomes $\bm{Y}(1)$ and $\bm{Y}(0)$ are randomly generated from an underlying data-generating process (Assumption~\ref{asp:iid}).
But conditioning on one realization $(\bm{Y}(1), \bm{Y}(0)) = (\bm{y}(1), \bm{y}(0))$, the causal effect $\tau_{\bm{y}(1), \bm{y}(0)}$ becomes deterministic.
Experimental designs that involve the above causal effect $\tau_{\bm{y}(1), \bm{y}(0)}$ are sometimes referred to as taking a ``design-based'' perspective, so named because the only source of randomness comes from the design of experiment.
Under the design-based perspective, we essentially wish to understand a finite sample.

The sampling-based and the design-based perspectives are highly related.
The average causal effect of the population and the average causal effect of a finite sample are connected through the following relationship,
\begin{align*}
\bE_\cF[\tau_{\bm{Y}(1), \bm{Y}(0)}] = \tau.
\end{align*}
This relationship is because of linearity of expectation and holds even without Assumption~\ref{asp:iid}.
For further discussions of the sampling-based and design-based perspectives, see \citet{abadie2020sampling, manski2018right}.

Now that we have prescribed the data-generating process, we move on to make assumptions on how to leverage the multiple units.
We introduce the following random assignment assumption.

\begin{assumption}[Random assignment]
\label{asp:RandomAssignment}
For each unit $j \in [n]$, the treatment assignment $W_j$ and the potential outcomes of all units $\big\{(Y_j(1), Y_j(0))\big\}_{j=1}^n$ are independent, that is,
\begin{align*}
\big\{(Y_j(1), Y_j(0))\big\}_{j=1}^n \independent W_j.
\end{align*}
\end{assumption}

Random assignment is an important property to have.
In fact, it is random assignment that warrants the literature of experimental design.
One easy way to satisfy the random assignment assumption is through conducting a ``randomized experiment.''
A randomized experiment $\eta: \{0,1\}^n \rightarrow[0,1]$ induces a joint discrete probability distribution over certain treatment assignment vectors $\bm{w}$, such that
\begin{align*}
\sum_{\bm{w} \in \{0,1\}^n} \eta(\bm{w}) = 1, && \eta(\bm{w}) \geq 0, \ \forall \ \bm{w} \in \{0,1\}^n. 
\end{align*}
We refer to such a joint discrete probability distribution $\eta$ as a design of experiment.
The treatment assignment vector $\bm{W}$ in this experiment conforms to the discrete probability distribution $\eta$; that is, $\Pr(\bm{W} = \bm{w}) = \eta(\bm{w})$.

Assumption~\ref{asp:RandomAssignment} means that $\eta$ must not depend on the potential outcomes of any unit.
Violation of Assumption~\ref{asp:RandomAssignment} leads to many active research directions in the literature; one of these is adaptive experiments.
See Section~\ref{sec:Survey} for further discussions.

Several traditional designs of randomized experiments satisfy Assumption~\ref{asp:RandomAssignment}, the random assignment assumption.
% Before we introduce a notion called covariates in Section~\ref{sec:Uncertainty2}, 
We now introduce two of the simplest designs: the Bernoulli design, and the completely randomized design.
Let $\bI\{w_j=1\}$ be an indicator function that takes the value of $1$ when $w_j=1$ and takes the value of $0$ otherwise. 

\begin{definition}[Bernoulli Design]
\label{defn:BD}
Under the Bernoulli design parameterized by treatment probability $p \in (0,1)$, the treatment assignment vector $\bm{W}$ is a vector of discrete random variables whose joint probability mass function is given by
% \begin{align}
% \eta(\bm{w}) = \left\{
% \begin{aligned}
% & \frac{ \prod_{j=1}^n p^{\bI\{w_j=1\}} (1-p)^{\bI\{w_j=0\}} }{1-p^n-(1-p)^n}, & & \ \text{if} \ \sum_{j=1}^n \bI\{w_j=1\} \ne 0, n \\
% & 0, & & o.w.
% \end{aligned} \right. \label{eqn:defn:BD}
% \end{align}
\begin{align}
\eta(\bm{w}) = \prod_{j=1}^n p^{\bI\{w_j=1\}} (1-p)^{\bI\{w_j=0\}}. \label{eqn:defn:BD}
\end{align}
\end{definition}
Intuitively, in a Bernoulli design, we use independent coin flips to determine the treatment assignments of all the units.
% Because we do not wish to have all the units assigned to the treatment group or the control group, we trim these two extreme cases, and then normalize the remaining probability point masses.
%Trimming the discrete probability distribution leads to the deeper literature of re-randomization theory.

In Definition~\ref{defn:BD}, each unit has the same treatment probability.
As a more general case, each unit could have a different treatment probability that is unrelated to the potential outcomes.
For example, in a Bernoulli design, each unit $j \in [n]$ could have a different treatment probability $p_j$, and the joint probability mass function can be written as $\eta(\bm{w}) = \prod_{j=1}^n p_j^{\bI\{w_j=1\}} (1-p_j)^{\bI\{w_j=0\}}.$

In addition to the Bernoulli design, there is another simple design called the completely randomized design.
Let $\dbinom{n}{pn}$ be the binomial coefficient of choosing $pn$ elements from a total of $n$ elements.

\begin{definition}[Completely Randomized Design]
\label{defn:CRD}
Let $pn \in [n-1]$ be an integer.
Under the completely randomized design parameterized by treatment probability $p \in (0,1)$, the treatment assignment vector $\bm{W}$ is a vector of discrete random variables whose joint probability mass function is given by
\begin{align}
\eta(\bm{w}) = \left\{
\begin{aligned}
& \frac{1}{ \dbinom{n}{pn} }, & & \ \text{if} \ \sum_{j=1}^n \bI\{w_j = 1\} = pn, \\
& 0, & & \text{otherwise}.
\end{aligned} \right. \label{eqn:defn:CRD}
\end{align}
% If $pn$ is not an integer, we will need to first decide to round up or round down $pn$ and then apply equation~\eqref{eqn:defn:CRD}.
\end{definition}
Intuitively, a completely randomized design first fixes the number of treatment and control units, and then randomly shuffles the units to determine which units receive treatment.

Different from the Bernoulli design, in a completely randomized design, the treatment assignments are not independent.
The treatment assignments of different units are negatively correlated; that is, $\Cov(\bI\{W_i=1\}, \bI\{W_j=1\}) < 0$ for any $i \ne j \in [n]$.
But the marginal distributions of all the treatment assignments are the same; that is, $\Pr(W_i=1) = \Pr(W_j=1)$ for any $i \ne j \in [n]$.

In a randomized experiment, we define, for each unit $j \in [n]$, the ``propensity'' to be the marginal probability that this unit receives treatment $\Pr(W_j = 1)$.
A randomized experiment sometimes satisfies the following assumption on the treatment probability.

\begin{assumption}[Positivity]
\label{asp:Positivity}
For each unit $j\in[n]$, the treatment probability is strictly between $0$ and $1$, that is,
\begin{align*}
0 < \Pr(W_j=1) < 1.
\end{align*}
\end{assumption}

We can verify that in the Bernoulli design (Definition~\ref{defn:BD}) and completely randomized design (Definition~\ref{defn:CRD}) above, as long as $p \in (0,1)$, both designs satisfy Assumption~\ref{asp:Positivity}.

Once the randomized experiment $\eta$ is determined, we could sample one treatment assignment vector $\bm{w}$ from the distribution induced by $\eta$.
Following this treatment assignment vector $\bm{w}$ and after we run the experiment, we could collect the observations and use them to estimate the causal effect.
We introduce two of the simplest estimators: the difference-in-means (DM) estimator and the inverse propensity weighting (IPW) estimator.

\begin{definition}[Difference-in-Means Estimator]
\label{defn:DM}
The difference-in-means estimator is defined as 
\begin{align}
\widehat{\tau}^{DM} = \frac{\sum_{j=1}^n Y_j \bI\{W_j=1\}}{\sum_{j=1}^n \bI\{W_j=1\}} - \frac{\sum_{j=1}^n Y_j \bI\{W_j=0\}}{\sum_{j=1}^n \bI\{W_j=0\}}, \label{eqn:defn:DM}
\end{align}
which requires $\sum_{j=1}^n \bI\{W_j=1\} \ne 0$ and $\sum_{j=1}^n \bI\{W_j=1\} \ne 0$; otherwise it is not well defined.
\end{definition}

The difference-in-means estimator, as its name suggests, simply compares the difference between the two sample means of the treatment and control groups.
For notational simplicity, we sometimes also write $N(1) = \sum_{j=1}^n \bI\{W_j=1\}$ and $N(0) = \sum_{j=1}^n \bI\{W_j=0\}$.
The difference-in-means estimator $\widehat{\tau}^{DM}$ can then be written as
\begin{align*}
\widehat{\tau}^{DM} = \frac{1}{N(1)}\sum_{j=1}^n Y_j \bI\{W_j=1\} - \frac{1}{N(0)}\sum_{j=1}^n Y_j \bI\{W_j=0\}.
\end{align*}
% When we replace $\bm{W}$ by $\bm{w}$ to stand for one realization, we also write $N(1), N(0)$ as $n(1), n(0)$.

\begin{definition}[IPW Estimator]
\label{defn:IPW}
The IPW estimator is defined as 
\begin{align}
\widehat{\tau}^{IPW} = \frac{1}{n} \sum_{j=1}^n \frac{Y_j \bI\{W_j=1\}}{\Pr\{W_j=1\}} - \frac{1}{n} \sum_{j=1}^n \frac{Y_j \bI\{W_j=0\}}{\Pr\{W_j=0\}}, \label{eqn:defn:IPW}
\end{align}
which requires $\Pr\{W_j=1\} \ne 0$ and $\Pr\{W_j=0\} \ne 0$ for every $j \in [n]$; otherwise it is not well defined.
\end{definition}

The IPW estimator, as its name suggests, weighs each observed outcome $Y_j$ by its inverse propensity $\frac{1}{\Pr(W_j=1)}$.
When each unit has the same treatment probability, such as in the Bernoulli design (Definition~\ref{defn:BD}) and the completely randomized design (Definition~\ref{defn:CRD}), the IPW estimator $\widehat{\tau}^{IPW}$ can be written as
\begin{align*}
\widehat{\tau}^{IPW} = \frac{\sum_{j=1}^n Y_j \bI\{W_j=1\}}{n \Pr\{W_j=1\}} - \frac{\sum_{j=1}^n Y_j \bI\{W_j=0\}}{n \Pr\{W_j=0\}}.
\end{align*}
From this expression we can see that the two estimators only differ by the denominator. 
The IPW estimator can be seen as replacing the random quantity $\sum_{j=1}^n \bI\{W_j=w\}$ in the denominators of the difference-in-means estimator by their expectations $n \Pr\{W_j=w\}$, for $w \in \{0,1\}$, respectively.

For both the Bernoulli design and the completely randomized design, both the difference-in-means estimator and the IPW estimator (leading to a total of four combinations) are well known to be ``unbiased,'' that is, the expectation of the estimator is equal to the causal effect that we wish to estimate. 

\begin{lemma}[Unbiasedness]
\label{lem:Unbiasedness}
We make Assumption~\ref{asp:NonInterference} and show the following four combinations all lead to unbiased estimators.
\begin{enumerate}
\item Under the Bernoulli design (Definition~\ref{defn:BD}), the difference-in-means estimator (Definition~\ref{defn:DM}) conditioning on $N(1) \notin \{0,n\}$ (otherwise, it is not well defined) is unbiased in estimating the average causal effect of a finite sample, that is,
\begin{align*}
\bE_{\eta}\big[ \widehat{\tau}^{DM} \big\vert N(1) \notin \{0,n\} \big] = \tau_{\bm{Y}(1), \bm{Y}(0)}.
\end{align*}
\item Under Assumption~\ref{asp:Positivity} and under the Bernoulli design (Definition~\ref{defn:BD}), the IPW estimator (Definition~\ref{defn:IPW}) is unbiased in estimating the average causal effect of a finite sample, that is,
\begin{align*}
\bE_{\eta}\big[ \widehat{\tau}^{IPW} \big] = \tau_{\bm{Y}(1), \bm{Y}(0)}.
\end{align*}
\item Under the completely randomized design (Definition~\ref{defn:CRD}, which automatically ensures $N(1) \notin \{0,n\}$), the difference-in-means estimator (Definition~\ref{defn:DM}) is unbiased in estimating the average causal effect of a finite sample, that is,
\begin{align*}
\bE_{\eta}\big[ \widehat{\tau}^{DM} \big] = \tau_{\bm{Y}(1), \bm{Y}(0)}.
\end{align*}
\item Under the completely randomized design (Definition~\ref{defn:CRD}, which automatically ensures Assumption~\ref{asp:Positivity}), the IPW estimator (Definition~\ref{defn:IPW}) is unbiased in estimating the average causal effect of a finite sample, that is,
\begin{align*}
\bE_{\eta}\big[ \widehat{\tau}^{IPW} \big] = \tau_{\bm{Y}(1), \bm{Y}(0)}.
\end{align*}
\end{enumerate}
Additionally, if we make Assumption~\ref{asp:iid}, then taking the expectation over the data-generating process $\cF$ and using the fact that $\bE_\cF[\tau_{\bm{Y}(1), \bm{Y}(0)}] = \tau$, we can also estimate the average causal effect of population $\tau$ unbiasedly.
\end{lemma}

So far we have seen that, under each combination of the choice of design of experiment and the choice of estimator, we can always estimate the causal effect (both $\tau_{\bm{Y}(1), \bm{Y}(0)}$ and $\tau$) unbiasedly.
A natural question to ask is which combination is more desirable to use.
To answer this question from an experimental design perspective, we usually fix the choice of estimator and then choose the optimal design of experiment.
% Note that, it remains an interesting open question what can one say about an estimator-robust design of experiment.

To choose the optimal design of experiment, we adopt the decision-theoretic framework.
The decision-theoretic framework was initially introduced by Abraham Wald in the 1940s \citep{wolfowitz1952abraham}, and was first adopted by \citet{wu1981robustness} to cast experimental design problems as optimization problems.
Under the decision-theoretic framework, we define the following risk function,
\begin{align*}
r(\eta, \bm{y}(1), \bm{y}(0)) = \sum_{\bm{w} \in \{0,1\}^n} \eta(\bm{w}) L(\bm{w}, \bm{y}(1), \bm{y}(0)),
\end{align*}
where $L(\bm{w}, \bm{y}(1), \bm{y}(0))$ is the loss function.
One of the most common choices of loss function is the square loss,
\begin{align*}
L(\bm{w}, \bm{y}(1), \bm{y}(0)) = \big(\widehat{\tau}(\bm{w}, \bm{y}(1), \bm{y}(0)) - \tau_{\bm{y}(1), \bm{y}(0)}\big)^2.
\end{align*}
Here we write the estimator in the form of $\widehat{\tau}(\bm{w}, \bm{y}(1), \bm{y}(0))$ to emphasize the dependence on the realized treatment assignments $\bm{w}$ and the realized potential outcomes $\bm{y}(1)$ and $\bm{y}(0)$.
Depending on our causal effect of interest, we could also consider a different square loss by replacing the $\tau_{\bm{y}(1), \bm{y}(0)}$ with $\tau$ and taking expectation of the risk function over an additional source of randomness $\cF$.
The square loss reflects our focus on the quality of estimation.
Other common loss functions include penalties associated with false discoveries (a notion related to hypothesis testing), or some costs or losses of revenue if one uses the estimator to directly guide further decisions.

After the loss function is specified, we wish to solve the following minimization problem,
\begin{align}
\min_{\eta} & \ r(\eta, \bm{y}(1), \bm{y}(0)), \qquad & & \text{(deterministic optimization)}. \label{eqn:framework:DeterministicOpt}
\end{align}
This optimization problem depends on $(\bm{y}(1), \bm{y}(0)) \in \bR^{2n}$ (imposing Assumption~\ref{asp:NonInterference}) the $2n$ potential outcomes.
If one already knew the $2n$ potential outcomes, then one could choose the design optimally by solving the deterministic optimization problem defined above (although if the $2n$ potential outcomes were already known then one could directly report the causal effect without having to run an experiment).
But more often, the potential outcomes are unknown.
This is essentially a problem of decision making under uncertainty.
So we can use a set of different tools to formulate a variety of optimization problems, such as the following two examples,
\begin{align}
\min_{\eta} & \ \bE_{(\bm{Y}(1), \bm{Y}(0)) \sim \cF^n} \big[ r(\eta, \bm{Y}(1), \bm{Y}(0)) \big], & & \text{(stochastic optimization)}. \label{eqn:framework:StochasticOpt} \\
\min_{\eta} & \ \max_{(\bm{y}(1), \bm{y}(0)) \in \cY} \ r(\eta, \bm{y}(1), \bm{y}(0)), & & \text{(robust optimization)}. \label{eqn:framework:RobustOpt}
\end{align}
The first formulation is often referred to as a stochastic optimization problem in the optimization literature, where the underlying data-generating process $\cF^n$ is assumed to be known.
Here we use $\cF^n$ to stand for the joint probability distribution of $\cF$ over all $n$ pairs of potential outcomes $(Y_j(1), Y_j(0))$.
This formulation is also referred to as Bayes rule in the experimental design literature \citep{kasy2016experimenters, lindley1956measure, rubin1978bayesian}.
The second formulation is often referred to as a robust optimization problem in the optimization literature, where $\cY$, the range of potential outcomes, is assumed to be known.
This formulation is also referred to as the minimax rule in the experimental design literature \citep{savage1951theory, wu1981robustness}.

These different formulations reflect different ways to model the uncertainty governing the potential outcomes $\bm{y}(1)$ and $\bm{y}(0)$.
Depending on how we model the uncertainty, we can choose the appropriate framework from \eqref{eqn:framework:DeterministicOpt} -- \eqref{eqn:framework:RobustOpt}.
In Sections~\ref{sec:Uncertainty1} --~\ref{sec:Uncertainty3}, we provide examples to illustrate how to model the uncertainty of potential outcomes $\bm{y}(1)$ and $\bm{y}(0)$ in a design of experiment, and cast an experimental design problem as a well-defined optimization problem.

\section{Modeling Uncertainty I: Robust Optimization}
\label{sec:Uncertainty1}

As introduced in Section~\ref{sec:ExpDesign} formulations~\eqref{eqn:framework:DeterministicOpt}~--~\eqref{eqn:framework:RobustOpt}, three lines of literature model uncertainty in the potential outcomes from three different perspectives.

The first line of literature often models uncertainty from a robust perspective; that is, there is an adversarial nature that generates the values of the potential outcomes.
To restrict this adversarial nature, the potential outcomes will usually take values from a family of candidate values, which are collectively referred to as an ``uncertainty set.''
The optimal design will usually be randomized in order to achieve minimax optimality.
The second line of literature often models uncertainty from a stochastic perspective; that is, the data-generating process of the potential outcomes is assumed to be known, yet the realized values of the potential outcomes are unknown.
The optimal design will usually be deterministic as the data-generating process is known.
The third line of literature often models uncertainty using specific models; that is, specific models are used to describe the data-generating process, and the uncertainty can be well explained.
The optimal design will usually be deterministic as the data-generating process is known and uncertainty can be well explained.

In this section, we illustrate the first line of literature.
The other two lines of literature will be illustrated in Sections~\ref{sec:Uncertainty2} and~\ref{sec:Uncertainty3}.

This first line of literature dates back to the seminal work of \citet{wu1981robustness}, who established a deep connection between experimental design and optimization.
% This manuscript further develops on \citet{wu1981robustness} and illustrates how to cast any experimental design problem as an optimization problem.
Below we present a simplified special case of \citet{wu1981robustness} with only two versions of treatment: one treatment and one control.
We adopt the following additive model for the potential outcomes,
\begin{align}
Y_j(w) = \alpha_w + g_j + \epsilon_{jw}, \quad \forall j \in [n], w \in \{0,1\} \label{eqn:Wu1981Model}
\end{align}
where $\alpha_w$ is the effect of treatment $w=1$ or control $w=0$;
$g_j$ is the unit fixed effect, which we assume to be deterministic and unknown; 
and $\epsilon_{jw}$ is the random noise with zero mean and equal variances $\sigma^2$.
The assumption that the random noises have equal variances is sometimes also referred to as the ``homoscedasticity'' assumption.
We further assume that the random noises are independent across $j \in [n]$, but are not necessarily independent between $w \in \{0,1\}$.
The linear regression literature sometimes specifies a model in the form of $Y_j(w) = \alpha_w + g_j + \epsilon_j$, by implicitly assuming that $\epsilon_{j1} = \epsilon_{j0} = \epsilon_j$.
Such a specification is usually equivalent to specifying the additive model \eqref{eqn:Wu1981Model}, as the specification in \eqref{eqn:Wu1981Model} allows the random noises $\epsilon_{j1}$ and $\epsilon_{j0}$ to have arbitrary correlations.
More fundamentally, this is because each unit never receives the treatment and control at the same time, so we never interact with $\epsilon_{j1}$ and $\epsilon_{j0}$ at the same time.

Collect $\bm{g} = (g_1, g_2, ..., g_n)$ in a vector form.
We assume that $\bm{g} \in \cG \subseteq \bR^n$ can take any element from a bounded set $\cG$.
% , which is referred to as a neighborhood of model violations by \citet{wu1981robustness} as it reflects the neighborhood from which the unit fixed effects can take their values.
We next make an assumption on $\cG$.
Denote $\pi: [n] \to [n]$ to be a permutation of $n$ elements; that is, $\pi$ is a one-to-one mapping between its domain and its range.
Denote $\Pi$ to be the ``permutation group'' on $n$ units; that is, it consists of all the $n!$ many possible permutations.
Other groups defined on $n$ units, such as a rotation group, are beyond the scope of this manuscript.
We refer to \citet{good2013permutation} for more details.
With a little abuse of notation, whenever we apply a permutation $\pi$ to a length-$n$ vector, we permute the elements in the vector; that is, we reload $\pi$ such that $\pi(\bm{g}) = (g_{\pi^{-1}(1)}, g_{\pi^{-1}(2)}, ..., g_{\pi^{-1}(n)})$.
Using the above notations, we introduce the following assumption with respect to a permutation group.

\begin{assumption}[Permutation Invariance under Permutation Group]
\label{asp:PermutationInvariance}
Let $\Pi$ be the permutation group on $n$ units.
The set $\cG \subseteq \bR^n$ is invariant under the permutation group $\Pi$, that is, 
\begin{align*}
\text{if } \ \bm{g} \in \cG \ \text{ then } \ \pi(\bm{g}) \in \cG, \ \forall \pi \in \Pi.
\end{align*}
\end{assumption} 

\begin{example}[Permutation Invariance]
\label{exa:Permutations}
\begin{figure}[!htb]
\centering
\caption{An illustration of a permutation $\pi: [n] \to [n]$.}
\label{fig:Permutations}
\begin{tikzpicture}
% Define the center point
\coordinate (center) at (0,0);
% Set the radius of the circles
\def\radius{0.2cm}
\def\positions{2cm}
\def\diff{1.8cm}
% Draw the circles and label them
\foreach \i/\number in {1/1, 2/2, 3/3, 4/4, 5/5} {
    \draw[line width=0.75pt] (center) ++(360/5*\i:\positions) circle (\radius) node {\number};
}
% Calculate the coordinates of the centers of the circles
\coordinate (first) at ($(center) + (360/5*1:\diff)$);
\coordinate (second) at ($(center) + (360/5*2:\diff)$);
\coordinate (third) at ($(center) + (360/5*3:\diff)$);
\coordinate (fourth) at ($(center) + (360/5*4:\diff)$);
\coordinate (fifth) at ($(center) + (360/5*5:\diff)$);
% Draw arrows between the circles
\draw[->, >=stealth, line width=0.75pt] (first) -- (third);
\draw[->, >=stealth, line width=0.75pt] (third) -- (fourth);
\draw[->, >=stealth, line width=0.75pt] (fourth) -- (fifth);
\draw[->, >=stealth, line width=0.75pt] (fifth) -- (second);
\draw[->, >=stealth, line width=0.75pt] (second) -- (first);
\end{tikzpicture}
\floatfoot{\textit{Note:} In this figure, one arrow stands for the permutation of an element; for example, the arrow from $1$ to $3$ stands for $\pi(1) = 3$. Because a permutation is an one-to-one mapping, each number serves as the start and the end of exactly one arrow, respectively.}
\end{figure}
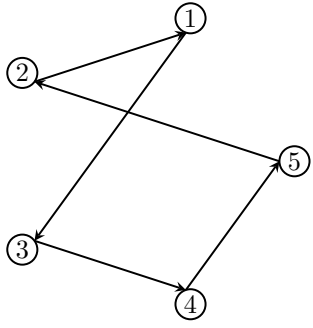
To illustrate the above permutation invariance assumption, consider a permutation $\pi \in \Pi$ from permutation group $\Pi$ such that $\pi(1) = 3, \pi(2) = 1, \pi(3) = 4, \pi(4) = 5, \pi(5) = 2.$ 
Then, we can correspondingly find $\pi^{-1}(1) = 2, \pi^{-1}(2) = 5, \pi^{-1}(3) = 1, \pi^{-1}(4) = 3, \pi^{-1}(5) = 4.$ 
See Figure~\ref{fig:Permutations} for an illustration.

Under the assumption that $\cG$ is invariant under the permutation group $\Pi$, and if we focus on permutation $\pi \in \Pi$ as defined above, then $\bm{g} = (2, 4, 6, 8, 10) \in \cG$ implies $\pi(\bm{g}) = (4, 10, 2, 6, 8) \in \cG$.
Furthermore, $\pi(\bm{g}) = (4, 10, 2, 6, 8) \in \cG$ implies $\pi(\pi(\bm{g})) = (10, 8, 4, 2, 6) \in \cG$, which further implies $(8, 6, 10, 4, 2) \in \cG$ and $(6, 2, 8, 10, 4) \in \cG$.
The permutation invariance assumption essentially assumes that the shape of the set $\cG$ is symmetric; that is, there exists a set of values $\cG_0 \subset \bR$ such that $\cG$ takes the form of $\cG = \cG_0^n$.
\end{example}

Under the model as defined in \eqref{eqn:Wu1981Model}, we consider the causal effect $\tau = \alpha_1 - \alpha_0$.
We also consider the simple difference-in-means estimator $\widehat{\tau} = \widehat{\tau}^{DM} = \frac{1}{N(1)}\sum_{j=1}^n Y_j \bI\{W_j=1\} - \frac{1}{N(0)}\sum_{j=1}^n Y_j \bI\{W_j=0\}.$
We then consider the loss function, for any realization of the treatment assignment vector $\bm{w}$ and any vector of unit fixed effects $\bm{g}$, to be
\begin{align*}
L(\bm{w}, \bm{g}) = \bE_{\bm{\epsilon}} \big[(\widehat{\tau} - \tau)^2\big]. %\label{eqn:Wu1981LossFunction}
\end{align*}
Here, because we adopt the model in \eqref{eqn:Wu1981Model}, we use $\bm{g}$ instead of $\bm{y}(1), \bm{y}(0)$ when writing out the loss function.
For any randomized design $\eta: \{0,1\}^n \to [0,1]$, recall the risk function is
\begin{align*}
r(\eta, \bm{g}) = \sum_{\bm{w} \in \{0,1\}^n} \eta(\bm{w}) L(\bm{w}, \bm{g}).
\end{align*}
We then adopt the robust optimization framework \eqref{eqn:framework:RobustOpt} as introduced at the end of Section~\ref{sec:ExpDesign} and consider the following minimax optimization problem
\begin{align*}
\min_{\eta} \max_{\bm{g} \in \cG} \ r(\eta, \bm{g}).
\end{align*}

To solve the above optimization problem, we make the following observation.
For any design $\eta$ and any permutation $\pi$, define $\eta_\pi$ to be a new design such that for any $\bm{w} \in \{0,1\}^n$, $\eta_\pi(\bm{w}) = \eta(\pi(\bm{w}))$.
Then, from any design $\eta$, we can construct a new design $\tilde{\eta}$ whose probability mass function is given as follows,
\begin{align*}
\tilde{\eta}(\bm{w}) = \frac{1}{n!} \sum_{\pi \in \Pi} \eta_\pi(\bm{w}), \ \forall \bm{w} \in \{0,1\}^n.
\end{align*}
Intuitively, $\tilde{\eta}$ is a design that assigns equal probability to each treatment assignment vector $\bm{w}$ and all its permuted vectors $\pi(\bm{w}), \forall \pi \in \Pi$.
More precisely, $\tilde{\eta}$ satisfies the property that $\tilde{\eta}(\bm{w}) = \tilde{\eta}(\pi(\bm{w}))$ for any $\pi \in \Pi$.
So $\tilde{\eta}$ can be interpreted as a distribution over completely randomized designs (Definition~\ref{defn:CRD}).
We next show that, under permutation invariance, the optimal design can be represented by a distribution over completely randomized designs.

% \begin{lemma}[\citet{wu1981robustness}. Optimality of Completely Randomized Designs under Permutation Invariance]
\begin{lemma}[\citet{wu1981robustness}]
\label{lem:CRDOptPermuInv}
Under the model considered in \eqref{eqn:Wu1981Model} (which implies Assumption~\ref{asp:NonInterference}) and under Assumption~\ref{asp:PermutationInvariance} if $\cG$ is permutation invariant, then for any $\eta$, 
\begin{align*}
\max_{\bm{g} \in \cG} \ r(\tilde{\eta}, \bm{g}) \ \leq \ \max_{\bm{g} \in \cG} \ r(\eta, \bm{g}).
\end{align*}
\end{lemma}

Now that we establish Lemma~\ref{lem:CRDOptPermuInv}, the only flexibility that we have in choosing the optimal design is to decide the sizes of the treatment and control groups.
Once the sizes are determined, they naturally determine the completely randomized design.
Then, to assign $n$ units to the treatment and control groups, we can choose to balance the sizes of both groups.
We formalize the above in the next result.

\begin{theorem}[\citet{wu1981robustness}. Optimal Design under Permutation Invariance]
\label{thm:CRDOptPermuInv}
Under the model considered in \eqref{eqn:Wu1981Model} (which implies Assumption~\ref{asp:NonInterference}) and under Assumption~\ref{asp:PermutationInvariance} if $\cG$ is permutation invariant, if $n$ is an even integer, and if we use the difference-in-means estimator to estimate the average treatment effect of the population,
then the balanced completely randomized design that randomly assigns $\frac{n}{2}$ units into the treatment group and the other $\frac{n}{2}$ units into the control group is minimax optimal with respect to $\cG$.
\end{theorem}

Permutation invariance is a powerful assumption to establish minimax optimality; \citet{bai2023randomize, basse2023minimax} have established similar results under permutation invariance.
As an alternative, below we introduce a different way of modeling uncertainty, still from a robust optimization perspective.
We adopt the robust optimization framework \eqref{eqn:framework:RobustOpt} as introduced at the end of Section~\ref{sec:ExpDesign} and directly model the potential outcomes $\bm{y}(1), \bm{y}(0)$ to come from a uniform uncertainty set.
We will introduce Lemma~\ref{lem:UniBSetReduction} and Theorem~\ref{thm:BDOptUniBSet}, which are probably too simple to have been studied in the existing literature.
Results in the same spirit have been shown in \citet{bojinov2023design, candogan2023correlated, ni2023design} yet they do not imply Lemma~\ref{lem:UniBSetReduction} and Theorem~\ref{thm:BDOptUniBSet}.

More specifically, we assume the existence of some positive constant $b>0$ such that the potential outcomes are bounded $Y_j(w) \in [-b, b], \forall j \in [n], w \in \{0,1\}$.
The range from which the potential outcomes can take values is given by $(\bm{y}(1), \bm{y}(0)) \in \cY = [-b, b]^{2n}$.
We consider the causal effect of $\tau_{\bm{y}(1), \bm{y}(0)}$ as defined in \eqref{eqn:tau:sample}.
We consider a Bernoulli design parameterized by probabilities $\bm{p} = (p_1, p_2, ..., p_n)$, where each unit $j \in [n]$ received treatment with probability $p_j$ and the treatment assignments across different units are independent.
We also consider the IPW estimator $\widehat{\tau} = \widehat{\tau}^{IPW}$ as defined in \eqref{eqn:defn:IPW}.
We then consider the loss function, for any realization of the treatment assignment vector $\bm{w}$, to be
\begin{align*}
L(\bm{w}, \bm{y}(1), \bm{y}(0)) = (\widehat{\tau} - \tau_{\bm{y}(1), \bm{y}(0)})^2.
\end{align*}
Note that, in Lemma~\ref{lem:CRDOptPermuInv} and Theorem~\ref{thm:CRDOptPermuInv}, we have focused on $\tau$. 
But here we focus on $\tau_{\bm{y}(1), \bm{y}(0)}$.
For any randomized design $\eta_{\bm{p}}: \{0,1\}^n \to [0,1]$, in which we use the subscript to emphasize the dependence on $\bm{p}$, the risk function can be expressed as
\begin{align*}
r(\eta_{\bm{p}}, \bm{y}(1), \bm{y}(0)) = \bE_{\bm{w} \sim \eta_{\bm{p}}}\big[L(\bm{w}, \bm{y}(1), \bm{y}(0))\big] = \bE_{\eta_{\bm{p}}}\big[(\widehat{\tau} - \tau_{\bm{y}(1), \bm{y}(0)})^2\big].
\end{align*}
We then adopt the robust optimization framework \eqref{eqn:framework:RobustOpt} as introduced at the end of Section~\ref{sec:ExpDesign} and consider the following minimax optimization problem,
\begin{align}
\min_{\eta_{\bm{p}}} \max_{(\bm{y}(1), \bm{y}(0)) \in \cY} \ r(\eta_{\bm{p}}, \bm{y}(1), \bm{y}(0)). \label{eqn:obj:UniBSet}
\end{align}

Through expanding the risk function, we can re-write the above optimization problem equivalently as follows.
\begin{lemma}
\label{lem:UniBSetReduction}
Under Assumption~\ref{asp:NonInterference}, if we conduct a Bernoulli design parameterized by probabilities $\bm{p} = (p_1, p_2, ..., p_n)$, the minimization problem as defined in \eqref{eqn:obj:UniBSet} is equivalent to the following minimization problem,
\begin{align*}
\min_{\eta_{\bm{p}}} \ \max_{(\bm{y}(1), \bm{y}(0)) \in \cY} \ \sum_{j=1}^n \frac{1}{p_j(1-p_j)} \Big( y_j(1) (1-p_j) + y_j(0) p_j \Big)^2.
\end{align*}
\end{lemma}

After this reduction, we can characterize the worst-case potential outcomes and then the vector of optimal treatment probabilities, which determines the optimal Bernoulli design.

\begin{theorem}[Optimal Bernoulli Design under Uniform Uncertainty Set]
\label{thm:BDOptUniBSet}
Under Assumption~\ref{asp:NonInterference}, if we conduct a Bernoulli design, if we consider a uniform uncertainty set $\cY = [-b, b]^{2n}$, and if we use the IPW estimator to estimate the average treatment effect of a finite sample,
then the balanced Bernoulli design that randomly assigns each unit into the treatment group and the control group with probability one half is minimax optimal with respect to $\cY$.
Moreover, the worst-case potential outcomes are such that $y_j(1) = y_j(0) = \pm b$.
\end{theorem}

Theorems~\ref{thm:CRDOptPermuInv} and~\ref{thm:BDOptUniBSet} present two examples of casting experimental design problems as robust optimization problems.
In both problems, no dominating strategy exists; that is, no design of experiment uniformly achieves the smallest loss over all the potential outcomes.
Nonetheless, we can first characterize the worst-case potential outcomes and then find the optimal design of experiment under such worst-case outcomes.

Generally speaking, there is usually no unifying approach in identifying the worst-case potential outcomes or in identifying the optimal design of experiment under the decision-theoretic framework in general.
It highly depends on the choice of the loss function, the choice of the causal effect and the estimator, the constraints on the design of experiment, and the framework we adopt in modeling the uncertainty behind the potential outcomes.
Theorems~\ref{thm:CRDOptPermuInv} and~\ref{thm:BDOptUniBSet} illustrate two combinations:
Theorem~\ref{thm:CRDOptPermuInv} uses the difference-in-means estimator to estimate the average treatment effect of the population, imposes no constraint on the design of experiment, and assumes permutation invariance;
Theorem~\ref{thm:BDOptUniBSet} uses the IPW estimator to estimate the average treatment effect of a finite sample, considers the family of Bernoulli designs, and assumes a uniform uncertainty set.
Other combinations can also be considered, yet the solution approach will likely be different.

The robust optimization framework has many variants.
Theorems~\ref{thm:CRDOptPermuInv} and~\ref{thm:BDOptUniBSet} are two examples of the basic minimax decision rule, which directly minimizes the worst-case risk as in \eqref{eqn:framework:RobustOpt}.
Other alternatives include the minimax regret decision rule \citep{manski2004statistical, stoye2009minimax} and the competitive analysis decision rule \citep{zhao2023adaptive}.

In this section, we have illustrated the robust optimization framework.
In Sections~\ref{sec:Uncertainty2} and~\ref{sec:Uncertainty3}, we will introduce the other two alternative frameworks of modeling uncertainty, which lead to stochastic and deterministic optimization problems, respectively.

\section{Modeling Uncertainty II: Stochastic Optimization}
\label{sec:Uncertainty2}

Recall that in Section~\ref{sec:Uncertainty1}, we have illustrated the first line of literature that models uncertainty under a robust optimization framework.
As a result, the optimal design is usually randomized to achieve minimax optimality.
In this section, we illustrate the second line of literature which models uncertainty under a stochastic optimization framework; that is, the data-generating process is usually given and known.
The optimal design aims at minimizing the risk function when uncertainty is governed by this data-generating process.
In contrast to the robust optimization framework, the optimal design under the stochastic optimization framework is usually deterministic, as the data-generating process is known.

The second line of literature dates back to at least the seminal work of \citet{lindley1956measure}, in the context of no covariates.
To better illustrate the stochastic optimization framework, we formally introduce covariates below.
For each unit $j\in[n]$, let there be an associated vector of covariates $\bm{X}_j$ that takes values from $\cX \subseteq \bR^d$.
We refer to $d$ as the ``dimension'' of the covariates.
Throughout this manuscript, we assume that dimension $d$ is much smaller than sample size $n$.
Such problems are often referred to as low-dimensional problems.
When dimension $d$ is comparable to, or even larger than, sample size $n$, it leads to high-dimensional problems.
We refer to \citet{buhlmann2011statistics, rigollet2019high, tibshirani1996regression, wainwright2019high} for further discussions.

We start by introducing an assumption on the underlying data-generating process of the covariates.
Similar to Assumption~\ref{asp:iid}, we assume that the covariates of different units are also generated from the same distribution.

\begin{assumption+}{\ref{asp:iid}$^*$}[Homogeneity]
\label{asp:iid-X}
For each unit $j \in [n]$, the potential outcomes and the covariates $(Y_j(1), Y_j(0), \bm{X}_j^\top)$ are independent and identical samples drawn from the same super-population, that is,
\begin{align*}
(Y_j(1), Y_j(0), \bm{X}_j^\top) \sim \cF.
\end{align*}
\end{assumption+}

To show the value of modeling covariates, we start from the following example in \citet{wager2020stats} about cash incentives for non-smoking.

\begin{example}[Simpson's Paradox]
\label{exa:SimpsonParadox}
Suppose we run two randomized experiments on giving teenagers cash incentives to discourage them from smoking.
These two randomized experiments are conducted in two locations, Geneva and Palo Alto, with a total of $\sim 18\%$ of the teenagers receiving treatment in Geneva and $\sim 6\%$ of the teenagers receiving treatment in Palo Alto.
See their respective summary statistics in Table~\ref{tbl:GenevaPaloAlto:Two}.

\begin{table}[h]
\caption{Summary statistics of Geneva and Palo Alto, respectively.}
\label{tbl:GenevaPaloAlto:Two}
\begin{minipage}{\textwidth}
\centering
\begin{tabular}{c|c|c|c}
Geneva      & Non-Smoker & Smoker & Smoke Rate \\ \hline
Treatment   & 581        & 350    & 37.59\%    \\ \hline
Control     & 2278       & 1979   & 46.49\%    \\ 
\end{tabular}
\end{minipage}
\begin{minipage}{\textwidth} 
\centering
\begin{tabular}{c|c|c|c}
Palo Alto   & Non-Smoker & Smoker & Smoke Rate \\ \hline
Treatment   & 152        & 5      & 3.18\%     \\ \hline
Control     & 2362       & 122    & 4.91\%     \\ 
\end{tabular}
\end{minipage}
\end{table}

If we look at each location separately, we can see that the treatment group has a lower smoke rate in each location.
However, if we combine two locations, we see a completely different result.
See the summary statistics in Table~\ref{tbl:GenevaPaloAlto:Combined}.

\begin{table}[h]
\caption{Summary statistics of combined data.}
\label{tbl:GenevaPaloAlto:Combined}
\centering
\begin{tabular}{c|c|c|c}
Geneva + Palo Alto & Non Smoker & Smoker & Smoke Rate \\ \hline
Treatment          & 733        & 355    & 32.63\%    \\ \hline
Control            & 4640       & 2101   & 31.17\%    \\ 
\end{tabular}
\end{table}

Although the data from each location suggest that the treatment group has a lower smoke rate, the combined data suggest that the treatment group has a higher smoke rate.
Such a counter-intuitive observation is referred to as Simpson's paradox.
\end{example}

If we carefully examine Example~\ref{exa:SimpsonParadox}, we will see that Assumption~\ref{asp:RandomAssignment} (i.e., the random assignment assumption) fails in the context of Simpson's paradox.
In the combined data, the potential outcomes are correlated with the treatment assignments because the treatment probabilities are different across the two locations, and the two locations also have different baseline smoking rates.
In this example, the location is referred to as a ``confounder.''

The potential existence of confounders motivates the conditional random assignment or the unconfoundedness assumption.

\begin{assumption+}{\ref{asp:RandomAssignment}$^*$}[Unconfoundedness]
\label{asp:Unconfoundedness}
For each unit $j \in [n]$, conditional on the covariates $\bm{X}_j$, the treatment assignment $W_j$ and the potential outcomes of all units $\big\{(Y_j(1), Y_j(0))\big\}_{j=1}^n$ are independent, that is,
\begin{align*}
\big\{(Y_j(1), Y_j(0))\big\}_{j=1}^n \independent W_j \vert \bm{X}_j.
\end{align*}
\end{assumption+}

In Example~\ref{exa:SimpsonParadox}, let $X_j$ be one single covariate indicating whether the experiment is conducted in Geneva or in Palo Alto.
Then, conditional on the location $X_j$, the treatment assignment is as good as random.
More generally, if the random treatment assignment only depends on covariates, we can generalize the definition of the propensity score from Section~\ref{sec:Intro} and define the propensity score as a function of covariates.
Define, for any unit $j \in [n]$ and conditional on its covariates taking values $\bm{X}_j = \bm{x}$, the propensity score $e(\bm{x})$ to be the probability that this unit receives treatment; that is, $e(\bm{x}) = \Pr(W_j = 1 \vert \bm{X}_j = \bm{x})$.
A randomized experiment sometimes satisfies the following assumption on the propensity score.

\begin{assumption+}{\ref{asp:Positivity}$^*$}[Positivity]
\label{asp:Positivity-X}
For any $\bm{x} \in \cX \subseteq \bR^d$, the propensity score is strictly between $0$ and $1$, that is,
\begin{align*}
0 < e(\bm{x}) < 1.
\end{align*}
\end{assumption+}

Simpson's Paradox in Example~\ref{exa:SimpsonParadox} and the definition of the propensity score motivate a design of experiment that depends on the covariates, which we refer to as a ``stratified randomized design.''

\begin{definition}[Stratified Randomized Design]
\label{defn:Stratified}
A stratified randomized experiment proceeds in the following two steps:
\begin{enumerate}
\item Form a partition of all the units $\cP = \{S_1, S_2, ..., S_k\}$, where we refer to each set $S_l$ as a stratum, such that 
\begin{align*}
\bigcup_{l=1}^k S_l = [n], \quad \text{ and } \quad S_l \cap S_{l'} = \emptyset, \ \forall \ l \ne l' \in [k].
\end{align*}
\item Within each stratum $S_l$, conduct a randomized design using any experimental design method, such that each unit from this stratum has a marginal probability $p_l$ of receiving treatment.
Two popular choices of the randomized design can be the Bernoulli design (Definition~\ref{defn:BD}) or the completely randomized design (Definition~\ref{defn:CRD}).
\end{enumerate}
\end{definition}

Each stratified randomized experiment is parameterized by a $k$-partition $\cP$ and a $k$-dimensional vector $\bm{p} \in (0,1)^k$.
In many applications, and similar to Example~\ref{exa:SimpsonParadox}, the treatment probabilities are different across different strata.
When the treatment probabilities are different across different strata, the traditional difference-in-means estimator $\widehat{\tau}^{DM}$ as defined in Definition~\ref{defn:DM} is no longer unbiased.
As an alternative, we aggregate the difference-in-means estimators from each stratum.

\begin{definition}[Aggregate Estimator]
\label{defn:Agg}
For each stratum $l \in [k]$, denote $s_l = |S_l|$. 
It is immediate to see that $\sum_{l=1}^k s_l = n$.
The aggregate estimator is defined as
\begin{align*}
\widehat{\tau}^{AGG} = \sum_{l=1}^k \frac{s_l}{n} \ \widehat{\tau}^{DM}_l,
\end{align*}
where
\begin{align*}
\widehat{\tau}^{DM}_l = \frac{\sum_{j \in S_l} Y_j \bI\{W_j=1\}}{\sum_{j \in S_l} \bI\{W_j=1\}} - \frac{\sum_{j \in S_l} Y_j \bI\{W_j=0\}}{\sum_{j \in S_l} \bI\{W_j=0\}}.
\end{align*}
\end{definition}

We can verify that, conditional on each stratum having at least one treatment unit and one control unit, the aggregate estimator $\widehat{\tau}^{AGG}$ is unbiased.
On the other hand, as long as Assumption~\ref{asp:Positivity-X} holds, the IPW estimator $\widehat{\tau}^{IPW}$ is also unbiased.
Both claims are similar to Lemma~\ref{lem:Unbiasedness} and their proofs follow similarly.

In some other applications, the treatment probabilities are the same across different strata; that is, $p_l = p, \forall l \in [k]$.
When the treatment probabilities are the same, the traditional difference-in-means estimator $\widehat{\tau}^{DM}$ is unbiased.
One simple special case is when we set $p_l = \frac{1}{2}, \ \forall l \in [k]$. 
Once $p_l = \frac{1}{2}, \ \forall l \in [k]$ is fixed, the key decision is then to choose the partition $\cP$.

Choosing $\cP$ can be done in different ways.
Next, we show how to choose $\cP$ from an optimization framework following \citet{bai2022optimality}.
A similar idea also appeared in \citet{kallus2018optimal}.
We make Assumption~\ref{asp:iid-X}, that is, the homogeneity assumption.
We then observe the covariates $\bm{x}_1, ..., \bm{x}_n$ and collect them into matrix $\bb{x} \in \bR^{n \times d}$ by stacking $\bm{x}_1^\top, ..., \bm{x}_n^\top$ by rows.
We would like to partition the units into different strata after observing their covariates.
We emphasize that the partition $\cP$ is after we observe all the covariates $\bb{x}$, and the optimal partition $\cP$ should depend on the observed covariates $\bb{x}$.
Yet we drop the dependence on $\bb{x}$ from $\cP$ when writing the partition.
We assume that the treatment probabilities are $\frac{1}{2}$ across all strata, and within each stratum we conduct a completely randomized experiment that is independent of any other stratum.

We consider the following average causal effect conditional on the covariates from a finite sample
\begin{align*}
\tau_{\bb{x}} = \frac{1}{n} \sum_{j=1}^n \bE_{(Y_j(1), Y_j(0)) \sim \cF \vert \bm{X}_j}\big[Y_j(1) - Y_j(0) \vert \bm{x}_j\big]
\end{align*}
conditional on all the covariates $\bb{x}$.
Here we use $\cF \vert \bm{X}_j$ to emphasize that conditional on the covariates being equal to $\bm{x}_j$, the potential outcomes of unit $j \in [n]$ may have a different distribution than unconditionally drawn from $\cF$.
So $\tau_{\bb{x}}$ may be different from $\tau = \bE[Y(1) - Y(0)]$ the causal effect of the underlying population.
But if we further take expectation then $\bE_{\bb{x} \sim \cF}[\tau_{\bb{x}}] = \tau$.

We then choose the estimator to estimate the causal effect $\tau_{\bb{x}}$.
Given the treatment probabilities are all equal to $\frac{1}{2}$ from all strata (so the difference-in-means estimator is unbiased), we consider the difference-in-means estimator $\widehat{\tau} = \widehat{\tau}^{DM} = \frac{1}{N(1)}\sum_{j=1}^n Y_j \bI\{W_j=1\} - \frac{1}{N(0)}\sum_{j=1}^n Y_j \bI\{W_j=0\}.$

We then consider the loss function, for any realization of the treatment assignment vector $\bm{w}$ and any realization of the potential outcome vectors $\bm{y}(1)$ and $\bm{y}(0)$, to be the square loss,
\begin{align*}
L(\bm{w}, \bm{y}(1), \bm{y}(0)) = (\widehat{\tau} - \tau_{\bb{x}})^2.
\end{align*}
For any stratified randomized design $\eta_{\cP}$, in which we use $\cP$ in the subscript to emphasize the dependence on the strata, and any realization of the potential outcome vectors $\bm{y}(1)$ and $\bm{y}(0)$, we write the risk function as
\begin{align*}
r(\eta_{\cP}, \bm{y}(1), \bm{y}(0)) = \bE_{\bm{W} \sim \eta_{\cP}} \big[ L(\bm{W}, \bm{y}(1), \bm{y}(0)) \big] = \bE_{\eta_{\cP}} \big[ (\widehat{\tau} - \tau_{\bb{x}})^2 \big].
\end{align*}
Finally, we adopt the stochastic optimization framework \eqref{eqn:framework:StochasticOpt} as introduced at the end of Section~\ref{sec:ExpDesign} and consider the following stochastic optimization problem,
\begin{align}
\min_{\cP} \ \bE_{(\bm{Y}(1), \bm{Y}(0)) \sim \cF^n \vert \bb{X}} \ \big[r(\eta_{\cP}, \bm{Y}(1), \bm{Y}(0) \vert \bb{x})\big] = \min_{\cP} \ \bE_{\eta_{\cP}, \cF^n \vert \bb{X}} \ \big[ (\widehat{\tau} - \tau_{\bb{x}})^2 \vert \bb{x} \big], \label{eqn:MatchedPairObj}
\end{align}
where we use $\cF^n \vert \bb{X}$ to stand for the joint probability distribution of $\cF$ over all the $n$ units, conditional on their covariates $\bb{X}$.
We use $\bE_{\eta_{\cP}, \cF^n \vert \bb{X}}\big[ (\widehat{\tau} - \tau_{\bb{x}})^2 \vert \bb{x} \big]$ to emphasize that randomness has two sources, one coming from the stratified randomization, and the other coming from the potential outcomes. 
Note that the randomness of potential outcomes is conditional on the covariates taking values $\bb{X} = \bb{x}$.
The decision that we make, $\cP$, should also depend on $\bb{x}$, but we make it implicit.

In what follows, we expand on the mean squared error as defined above in \eqref{eqn:MatchedPairObj}.
Before we proceed, we examine a simple result that holds generally.

\begin{lemma}[Bias-Variance Trade-off]
\label{lem:BiasVariance}
For any estimator $\widehat{\tau}$ and any causal effect $\tau$, the mean squared error of this estimator can be decomposed as
\begin{align*}
\bE\big[(\widehat{\tau}-\tau)^2\big] = & \ \bE\big[(\widehat{\tau}-\bE[\widehat{\tau}]+\bE[\widehat{\tau}]-\tau)^2\big] \\
= & \ \bE\big[(\widehat{\tau}-\bE[\widehat{\tau}])^2\big] + (\bE[\widehat{\tau}]-\tau)^2 + 2 \bE\big[\widehat{\tau}-\bE[\widehat{\tau}]\big] \cdot (\bE[\widehat{\tau}]-\tau) \\
= & \ \Var(\widehat{\tau}) + \Bias(\widehat{\tau})^2.
\end{align*}
\end{lemma}

Using Lemma~\ref{lem:BiasVariance}, the mean squared error of any estimator can be decomposed into two terms, its variance term, and the square of its bias term.
This decomposition usually allows for the study of the bias-variance trade-off.

Using results similar to Lemma~\ref{lem:BiasVariance}, we show that minimizing the mean squared error in the objective function of \eqref{eqn:MatchedPairObj} is equivalent to the following minimization problem.
\begin{lemma}[\citet{bai2022optimality}]
\label{lem:MatchedPairReduction}
Under Assumption~\ref{asp:NonInterference} and Assumption~\ref{asp:iid-X}, if $n$ is an even integer, if the treatment probabilities are $\frac{1}{2}$ across all strata, and if within each stratum we conduct an independent completely randomized experiment, the minimization problem as defined in \eqref{eqn:MatchedPairObj} is equivalent to the following minimization problem,
\begin{align*}
\min_{\eta_{\cP}} \ \Var_{\eta_{\cP}} \bigg( \sum_{j=1}^n \Big( \bE_{\cF \vert \bm{X}_j}[Y_j(1)+Y_j(0)\vert \bm{x}_j] \bI\{W_j=1\} \Big) \bigg).
\end{align*}
\end{lemma}

Now we define
\begin{align*}
g(\bm{x}) = \bE_{(Y_j(1), Y_j(0)) \sim \cF \vert \bm{X}_j} \big[ Y_j(1) + Y_j(0) \vert \bm{x}\big]
\end{align*}
to be the baseline function.
In Example~\ref{exa:SimpsonParadox}, this function stands for the baseline smoke rate in Geneva or Palo Alto.
For each unit $j \in [n]$, let the baseline be $g_j = g(\bm{x}_j)$ and collect all the baselines in a vector form as $\bm{g} = (g_1, g_2, ..., g_n)$.
If the baselines are known and given, we can use the baselines to re-write the expression in Lemma~\ref{lem:MatchedPairReduction} as
\begin{align}
\min_{\eta_\cP} \ \bm{g}^\top \bb{v} \bm{g}, \label{eqn:QPVarCov}
\end{align}
where $\bb{v}$ is the following covariance matrix 
\begin{align*}
% \bb{v} = 
\left[
\footnotesize
\begin{array}{cccc} 
\Var_{\eta_\cP}(\bI\{W_1=1\}) & \Cov_{\eta_\cP}(\bI\{W_1=1\}, \bI\{W_2=1\}) & \ldots & \Cov_{\eta_\cP}(\bI\{W_1=1\}, \bI\{W_n=1\}) \\ 
\Cov_{\eta_\cP}(\bI\{W_2=1\}, \bI\{W_1=1\}) & \Var_{\eta_\cP}(\bI\{W_2=1\}) & \ldots & \Cov_{\eta_\cP}(\bI\{W_2=1\}, \bI\{W_n=1\}) \\ 
\vdots & \vdots & & \vdots \\
\Cov_{\eta_\cP}(\bI\{W_n=1\}, \bI\{W_1=1\}) & \Cov_{\eta_\cP}(\bI\{W_n=1\}, \bI\{W_2=1\}) & ... & \Var_{\eta_\cP}(\bI\{W_n=1\}) \\
\end{array}\right].
\end{align*}

Because the randomization is independent across different strata, the covariance between the treatment assignments of any two units across two different strata is zero.
So the covariance matrix $\bb{v}$ can be decomposed into several diagonal blocks, with the off-diagonal components being equal to zero.
We can then look into the diagonal blocks and show that the optimal partition consists of only size-two strata.

\begin{theorem}[\citet{bai2022optimality}. Optimal Matched Pair Design]
\label{thm:MatchedPairOpt}
Without loss of generality re-arrange the baselines from high to low such that $g_1 \geq g_2 \geq ... \geq g_n$.
Under Assumption~\ref{asp:NonInterference} and Assumption~\ref{asp:iid-X}, if $n$ is an even integer, if the treatment probabilities are $\frac{1}{2}$ across all strata, and if within each stratum we conduct an independent completely randomized experiment, the optimal solution to the minimization problem as defined in \eqref{eqn:MatchedPairObj} is such that
\begin{align*}
\cP = \Big\{ \{1,2\}, \{3,4\}, ..., \{n-1, n\} \Big\}.
\end{align*}
\end{theorem}

Theorem~\ref{thm:MatchedPairOpt} presents an example of how to cast a covariate-dependent experimental design problem as a stochastic optimization problem.
Usually, the optimal solution to a stochastic optimization problem is deterministic unless the problem has multiple optimal solutions.
As we have seen from Theorem~\ref{thm:MatchedPairOpt}, because our decision space is over the stratification, the optimal stratification is deterministic.
However, in a stratified randomized experiment, we conduct a randomized experiment within each stratum.
Even though the optimal decision (i.e., stratification) is deterministic, the experiment itself is still randomized as a result of randomization within each stratum.

The optimal matched pair design as in Theorem~\ref{thm:MatchedPairOpt} is a special case of the stochastic optimization framework.
The optimal design only requires knowledge about the baseline function $g(\cdot)$, which only depends on the first moments of the joint distribution.
The optimal design does not require knowledge about the entire joint distribution $(Y_j(1), Y_j(0)) \sim \cF \vert \bm{X}_j$.
Generally speaking, however, the joint distribution will probably be required to find the optimal design under the stochastic optimization framework.

To conclude this section, we point out that we have used the baseline function $g(\cdot)$ to guide the randomization.
In Section~\ref{sec:Uncertainty3} we will take a deeper look at the baseline function $g(\cdot)$ by imposing a linear model, which sometimes leads to deterministic optimization problems.

% \subsection*{Bibliographical Notes}
% There are different ways to choose $\cP$.
% Another mathematically elegant way of choosing $\cP$ is to use threshold blocking [Higgins, Savje, Sekhon 2016 PNAS].
% Let $c_{ij}, \forall i,j\in[n]$ be the distance between units $i$ and $j$.
% The authors aim at finding 
% \begin{align*}
% \min_{\cP} \ \max_{i,j \in S, S\in \cP} \ c_{ij}.
% \end{align*}
% Under triangular assumption, that is, $c_{ij} + c_{jk} > c_{ik}, \forall i,j,k \in [n]$, there exists a $2$-approximation algorithm in finding a near-optimal partition $\cP$.

\section{Modeling Uncertainty III: Deterministic Optimization}
\label{sec:Uncertainty3}

Recall that in Section~\ref{sec:Uncertainty1}, we have illustrated the first line of literature that models uncertainty under a robust optimization framework; in Section~\ref{sec:Uncertainty2}, we have illustrated the second line of literature that models uncertainty under a stochastic optimization framework.
In this section, we illustrate the third line of literature, which uses specific models to explain uncertainty.
As uncertainty can be well explained, the optimal design is usually deterministic.

The third line of literature dates back to the seminal work of \citet{smith1918standard}.
Following the notations in \citet{sibson1974optimality}, we adopt a linear additive model that is similar to \eqref{eqn:Wu1981Model} as described in Section~\ref{sec:Uncertainty1} to incorporate covariates,
\begin{align}
Y_j(w) = \tau w + \bm{\beta}^\top \bm{X}_j + \epsilon_{jw}, \quad \forall j \in [n], w \in \{0,1\}, \label{eqn:LinearModel}
\end{align}
where $\tau$ is the causal effect that we are interested in, which corresponds to $(\alpha_1 - \alpha_0)$ in model \eqref{eqn:Wu1981Model};
$\bm{X}_j$ is a vector of covariates of unit $j$; 
$\bm{\beta}$ is a vector of unknown parameters;
and $\epsilon_{jw}$ is the random noise with zero mean and equal variance $\sigma^2$.
The random noises are independent across $j \in [n]$, but are not necessarily independent between $w \in \{0,1\}$.
We do not specify a fixed effect here, because for any unit $j\in[n]$, we can always augment the vector $\bm{X}_j$ by adding one extra dimension to $\bm{X}_j$ such that the first dimension of $\bm{X}_j$ is equal to $1$.
Such an augmentation allows $\bm{\beta}^\top\bm{X}_j$ to incorporate a non-zero intercept.

Under the linear additive model \eqref{eqn:LinearModel}, we usually use linear regression to estimate the causal effect $\tau$ and the coefficients $\bm{\beta}$.
To succinctly describe the linear regression, we introduce the following notations.
We collect causal effect $\tau$ and coefficients $\bm{\beta}$ into a vector and denote $\bm{\theta} = (\tau, \bm{\beta}^\top)^\top \in \bR^{d+1}$.
We collect all covariates into a matrix $\bb{X} \in \bR^{n \times d}$ by stacking $\bm{X}_1^\top, ..., \bm{X}_n^\top$ by rows.
For any random treatment assignment vector $\bm{W}$ (which possibly depends on the covariates $\bb{X}$), we collect them with $\bb{X}$ and denote $\bb{Z} = \big[ \bm{W} \ \bb{X} \big]$, which takes values from $\{0,1\}^{n \times 1} \times \bR^{n \times d}$.
We also collect all observed outcomes $Y_j, \forall j \in [n]$ into a vector $\bm{Y} \in \bR^n$.
Let $\epsilon_{j1} = \epsilon_{j0}, \forall j \in [n]$ and denote $\epsilon_j = \epsilon_{j1} = \epsilon_{j0}, \forall j \in [n]$.
We collect all random noises $\epsilon_{j}, \forall j \in [n]$ into a vector $\bm{\epsilon}$ that takes values from $\bR^n$.
We then have \eqref{eqn:LinearModel} on the $n$ samples,
\begin{align*}
\bm{Y} = \bb{Z} \bm{\theta} + \bm{\epsilon}.
\end{align*}
Under the homoscedasticity assumption, we usually estimate $\bm{\theta}$ using the ordinary least squares (OLS) estimator, which is defined as follows.

\begin{definition}[OLS Estimator]
The OLS estimator is defined as
\begin{align*}
\widehat{\bm{\theta}} = (\bb{Z}^\top \bb{Z})^{-1} \bb{Z}^\top \bm{Y}.
\end{align*}
The OLS estimator can be seen as the minimizer to the following unconstrained quadratic program,
\begin{align*}
\widehat{\bm{\theta}} = \min_{\bm{\theta}} \big\| \bm{Y} - \bb{Z} \bm{\theta} \big\|^2_2,
\end{align*}
where $\| \cdot \|_2$ stands for the $L_2$ norm of a vector.
\end{definition}

It is well-known that the OLS estimator is unbiased. 
To see this, we re-write the OLS estimator as
\begin{align*}
\widehat{\bm{\theta}} = (\bb{Z}^\top \bb{Z})^{-1} \bb{Z}^\top \bm{Y} = (\bb{Z}^\top \bb{Z})^{-1} \bb{Z}^\top (\bb{Z} \bm{\theta} + \bm{\epsilon}) = \bm{\theta} + (\bb{Z}^\top \bb{Z})^{-1} \bb{Z}^\top \bm{\epsilon}.
\end{align*}
Because the noises $\epsilon_{j}, \forall j \in [n]$ all have zero means, the OLS estimator is unbiased, that is,
\begin{align}
\bE_{\bm{\epsilon}}\big[\widehat{\bm{\theta}}\big] - \bm{\theta} = \bE_{\bm{\epsilon}}\big[(\bb{Z}^\top \bb{Z})^{-1} \bb{Z}^\top \bm{\epsilon}\big] = 0. \label{eqn:OLS:Unbiased}
\end{align}
The source of randomness behind this unbiasedness comes from the random noises $\bm{\epsilon}$.
Regardless of the design of experiment (as long as $\bb{Z}^\top \bb{Z}$ is invertible), this unbiasedness result holds.

Below we study how to choose a design of experiment when we use the OLS estimator.
Suppose we are given $\bm{x}_1, ..., \bm{x}_n$, the realizations of the covariates from the $n$ units.
We now wish to assign all $n$ units into treatment and control groups.
To do so, we start with one simple objective.

Under the model as defined in \eqref{eqn:LinearModel}, we consider the causal effect $\tau$.
We also consider the OLS estimator $\widehat{\tau} = \widehat{\tau}^{OLS}$, which is the first dimension of the estimator $\widehat{\bm{\theta}}$.
We then consider the loss function, for any realization of the treatment assignment vector $\bm{w}$, to be
\begin{align*}
L(\bm{w}) = \bE_{\bm{\epsilon}} \big[(\widehat{\tau} - \tau)^2\big] = \Var_{\bm{\epsilon}}\big( \widehat{\tau} \big). 
\end{align*}
The second equality holds because of expression \eqref{eqn:OLS:Unbiased}; that is, the OLS estimator is unbiased.
Note that as the loss function is deterministic, the optimal design $\eta: \{0,1\}^n \to [0,1]$ must also be deterministic; that is, the optimal design must put all probability mass on one support unless the problem has multiple optimal solutions.
So the risk minimization problem is equivalent to the following deterministic optimization problem,
\begin{align}
\min_{\eta} \ r(\eta) = \min_{\bm{w} \in \{0,1\}^n} \ \Var_{\bm{\epsilon}}\big( \widehat{\tau} \big). \label{eqn:D_A:OptimalDesign}
\end{align}
Problem \eqref{eqn:D_A:OptimalDesign} above is sometimes referred to as the $D_A$-optimal experimental design problem.

To solve the $D_A$-optimal experimental design problem, we examine the covariance matrix of the OLS estimator.
Below we write $\bb{z}$ and $\bb{x}$ to emphasize that we are given the realizations of the covariates $\bm{x}_1,...,\bm{x}_n$.
Note that
\begin{align}
\Var_{\bm{\epsilon}}\big(\widehat{\bm{\theta}}\big) = (\bb{z}^\top \bb{z})^{-1} \bb{z}^\top \Var_{\bm{\epsilon}}(\bm{\epsilon}\bm{\epsilon}^\top) \bb{z} (\bb{z}^\top \bb{z})^{-1} = \sigma^2 (\bb{z}^\top \bb{z})^{-1}, \label{eqn:ErrorCovarianceMatrix}
\end{align}
where the second equality is because $\Var_{\bm{\epsilon}}(\bm{\epsilon}\bm{\epsilon}^\top)$ is an $n \times n$ diagonal matrix with diagonal elements equal to $\sigma^2$ (assuming homoscedasticity) and the off-diagonal elements equal to $0$.
The variance of $\widehat{\tau}$ (the first element of $\widehat{\bm{\theta}}$) is then equal to the first element in the first row of the matrix $\Var_{\bm{\epsilon}}\big(\widehat{\bm{\theta}}\big)$, that is, $\sigma^2 \bm{e}_1^\top (\bb{z}^\top \bb{z})^{-1} \bm{e}_1$, where $\bm{e}_1 = (1, 0, 0, ..., 0)^\top$ stands for a basis vector with only a one as the first element and zero otherwise.
Using the bordering method in block matrix inversion, we can calculate $(\bb{z}^\top \bb{z})^{-1}$ and obtain
\begin{align*}
\Var_{\bm{\epsilon}}(\widehat{\tau}) = \frac{\sigma^2}{\bm{w}^\top \bm{w} - \bm{w}^\top \bb{x} (\bb{x}^\top \bb{x})^{-1} \bb{x}^\top \bm{w}}.
\end{align*}

\begin{lemma}[\citet{sibson1974optimality, bhat2020near}. $D_A$-Optimal Design]
Let $\bb{I}_n$ be the $n \times n$ identity matrix.
The $D_A$-optimal design problem is equivalent to maximizing the following problem of 
\begin{align*}
\max_{\bm{w} \in \{0,1\}^n} \bm{w}^\top (\bb{I}_n - \bb{x} (\bb{x}^\top \bb{x})^{-1} \bb{x}^\top) \bm{w}.
\end{align*}
\end{lemma}

Many algorithms have been used to (approximately) solve the $D_A$-optimal design problem.
One algorithm, given by \citet{bhat2020near}, is to reduce this problem to the maximum cut problem, which converts the \citet{goemans1995improved} result into a $\frac{2}{\pi}$-approximate solution to this $D_A$-optimal design problem.
We skip the details of the approximation algorithm as they go beyond the scope of this manuscript.

The optimal experimental design problem has a different version, and this different version is probably more widely studied in the literature.
In the above derivation, we allocate all the $n$ units into either the treatment or the control group.
We then use all the data from these $n$ units to construct the OLS estimator to estimate the causal effect $\tau$.
In the different version of the optimal experimental design problem, the decision is to select a subset of the $n$ units to conduct experiments, and only one version of treatment (rather than treatment and control) is involved in the experiment.
Because there is only one version of treatment, the causal effect $\tau$ is not well-defined, and the only focus is on estimating the unknown parameters $\bm{\beta}$.

Mathematically, letting $\bm{x}_j$ be the covariates for unit $j \in [n]$, the decision is to choose a subset $S \subseteq [n]$, with at most cardinality $|S| \leq k$, to construct the OLS estimator,
\begin{align*}
\widehat{\bm{\beta}} = \Big(\sum_{j \in S} \bm{x}_j \bm{x}_j^\top\Big)^{-1} \sum_{j \in S} \bm{x}_j Y_j.
\end{align*}
The objective is related to the covariance matrix of $\Var_{\bm{\epsilon}}\big(\widehat{\bm{\beta}}\big)$.
The $D_A$-optimal design problem is concerned with minimizing a scalarization of the covariance matrix $\Var_{\bm{\epsilon}}(\widehat{\bm{\beta}})$ on some dimension $l \in [d]$.
Many other objectives can be considered. 
Among them, one of the most popular is the $D$-optimal criterion \citep{fedorov2013theory, pukelsheim2006optimal, silvey2013optimal}.
Recall from \eqref{eqn:ErrorCovarianceMatrix} that
\begin{align*}
\Var_{\bm{\epsilon}}\big(\widehat{\bm{\beta}}\big) = \sigma^2 \Big(\sum_{j \in S} \bm{x}_j \bm{x}_j^\top\Big)^{-1}.
\end{align*}
The $D$-optimal design minimizes the determinant of this covariance matrix, where the letter $D$ stands for ``determinant.''
The $D$-optimal design also has a geometric interpretation of minimizing the volume of an ellipsoid at any fixed confidence level \citep{titterington1975optimal}.
Using the fact that the determinant of the inverse of an invertible matrix is the reciprocal of the determinant of the matrix, that is, $\det(\bb{A}^{-1}) = \det(\bb{A})^{-1}$, the $D$-optimal design is equivalent to minimizing
\begin{align*}
\min_{S \subseteq [n]} \det\bigg(\sum_{j \in S} \bm{x}_j \bm{x}_j^\top\bigg)^{-\frac{1}{D}}.
\end{align*}
This optimization problem is computationally challenging, and many efforts have been made to computationally solve this problem. 
We refer to \citet{allen2021near, madan2019combinatorial, meyer1995coordinate, nikolov2015randomized, nikolov2016maximizing, singh2018approximate, singh2020approximation, summa2014largest} for recent developments.

So far we have seen the covariates that are directly observed.
At times, we may also worry about the unobserved covariates.
One way to incorporate the unobserved covariates is through historical data, using the synthetic control method \citep{abadie2021using, abadie2010synthetic, abadie2003synthetic}.
To introduce the synthetic control method, we first introduce the panel data.
Let there be $n$ units, denoted as $[n]$.
We focus on a setting where these $n$ units are fixed (instead of randomly sampled from a super-population).
Let there be $T$ discrete, finite time periods, denoted as $[T]$.
In the panel data setting, the convention is to use the capital letter $T$ for a fixed time horizon; even though we use the capital letter $T$, it is still a constant, not a random variable.
In panel data, each unit $j \in [n]$ is repeatedly exposed to treatment or control, as well as repeatedly observed for a duration of time.
For each unit $j \in [n]$ and at each time period $t \in [T]$, we denote the treatment assignment as $W_{jt}$, which takes values from $\{0,1\}$, and the observed outcome as $Y_{jt}$, which takes values from $\bR$.
The observed outcomes are connected to the potential outcomes by 
\begin{align*}
Y_{jt} = \left\{ 
\begin{aligned}
Y_{jt}(1), & \ \text{if} \ W_{jt} = 1, \\
Y_{jt}(0), & \ \text{if} \ W_{jt} = 0.
\end{aligned}
\right.
\end{align*}

Suppose we stand at the end of time period $T_0$.
We refer to such $T_0$ periods as the pre-experimental periods.
The last $(T-T_0)$ periods $\{T_0+1, ..., T\}$ are the experimental periods.
All units receive control during the pre-experimental periods; that is, $W_{jt} = 0, \forall j \in [n], t \in [T_0]$.
At the end of time period $T_0$ and after collecting the observed outcomes up until period $T_0$, we choose a subset of the $n$ units to receive treatment during the experimental periods, leaving the remaining units in control.

\begin{example}
See Table~\ref{tbl:PanelData} for an illustration of the panel data. 
In Table~\ref{tbl:PanelData}, units~1 and~2 are the only two treatment units, and the gray area indicates that these two units receive treatment.
\begin{table}[htb]
\centering
\begin{tabular}{c|ccccccc}
            & Time 1 & Time 2 & $\ldots$ & Time $T_0$ & Time $(T_0+1)$                                  & $\ldots$                                        & Time $T$                                        \\ \hline
Unit $1$ &        &        &          &            & \cellcolor[HTML]{9B9B9B}{\color[HTML]{9B9B9B} } & \cellcolor[HTML]{9B9B9B}{\color[HTML]{9B9B9B} } & \cellcolor[HTML]{9B9B9B}{\color[HTML]{9B9B9B} } \\
Unit $2$ &        &        &          &            & \cellcolor[HTML]{9B9B9B}{\color[HTML]{9B9B9B} } & \cellcolor[HTML]{9B9B9B}{\color[HTML]{9B9B9B} } & \cellcolor[HTML]{9B9B9B}{\color[HTML]{9B9B9B} } \\
Unit $3$ &        &        &          &            &                                                 &                                                 &                                                 \\
$\vdots$    &        &        &          & $Y_{jt}$   &                                                 &                                                 &                                                 \\
Unit $n$ &        &        &          &            &                                                 &                                                 &                                                
\end{tabular}
\caption{An illustration of panel data}
\label{tbl:PanelData}
\end{table}
\end{example}

We adopt a linear model to incorporate both observed covariates and unobserved covariates.
Such a model is often referred to as a linear factor model.
\begin{align}
Y_{jt}(w) = \alpha_t(w) + \bm{\beta}_t(w)^\top \bm{X}_j + \bm{\lambda}_t(w)^\top \bm{\mu}_j + \epsilon_{jt}(w), \quad \forall j \in [n], t \in [T], w \in \{0,1\}, \label{eqn:LinearFactorModel}
\end{align}
where $\alpha_t(w)$ stands for a fixed effect;
$\bm{X}_j$ stands for a column vector of observed covariates that takes values from $\bR^d$;
$\bm{\mu}_j$ stands for a column vector of unobserved covariates that takes values from $\bR^r$;
$\bm{\beta}_t(w) \in \bR^d$ and $\bm{\lambda}_t(w) \in \bR^r$ stand for two column vectors of unknown parameters;
and $\epsilon_{jt}(w)$ is the random noise with zero mean and sub-Gaussian tail whose variance proxy is upper bounded by $\overline{\sigma}^2$.
The random noises are independent across $j \in [n]$, but are not necessarily independent between $w \in \{0,1\}$.

We use the synthetic control method to choose the treatment and control units, and to estimate the causal effect.
For each unit $j \in [n]$ and at any time period $t \in [T]$, let $\tau_{jt} = Y_{jt}(1) - Y_{jt}(0)$ be the individual causal effect.
We are given a vector of constants $\bm{f} = (f_1, ..., f_n) \in [0,1]^n$, such that $\sum_{j=1}^n f_j = 1$.
We are interested in the following average treatment effect for any $t \in \{T_0+1,...,T\}$,
\begin{align*}
\tau_t = \sum_{j=1}^n f_j \tau_{jt} = \sum_{j=1}^n f_j \big(Y_{jt}(1) - Y_{jt}(0)\big).
\end{align*}
In this setting, the synthetic control estimator takes as inputs two vectors of weights, $\bm{u} = (u_1, ..., u_n), \bm{v} = (v_1, ..., v_n)\in [0,1]^n$, such that $\sum_{j=1}^n u_j = \sum_{j=1}^n v_j = 1$ and $u_j v_j = 0, \forall j \in [n]$.
Once $\bm{u}$ and $\bm{v}$ are determined, the synthetic control estimator for any $t \in \{T_0+1,...,T\}$ is given by
\begin{align*}
\widehat{\tau}^{SC}_t(\bm{u}, \bm{v}) = \sum_{j=1}^n u_j Y_{jt} - \sum_{j=1}^n v_j Y_{jt}.
\end{align*}
Intuitively, the synthetic control estimator aims at constructing a synthetic treatment unit $\sum_{j=1}^n u_j Y_{jt}$ that mimics the unobservable averaged treatment outcome $\sum_{j=1}^n f_j Y_{jt}(1)$, as well as a synthetic control unit $\sum_{j=1}^n v_j Y_{jt}$ that mimics the unobservable averaged control outcome $\sum_{j=1}^n f_j Y_{jt}(0)$.
In observational studies, we usually only construct the synthetic control unit, and this is where the name ``synthetic control'' comes from.

We next introduce a quadratic program to determine the weights $\bm{u}$ and $\bm{v}$.
Let $k \leq n$ be a small positive integer that restricts the number of treatment units.
For any $j \in [n]$, let $\bm{Y}_j = (Y_{j1}, Y_{j2}, ..., Y_{jT_0})$ be the vector of observed outcomes during the pre-experimental periods. 
We then denote $\bm{Z}_j = (\bm{Y}_j^\top, \bm{X}_j^\top)^\top$ and $\overline{\bm{Z}} = \sum_{j=1}^n f_j \bm{Z}_j$.
At the end of period $T_0$, conditional on $\bm{Y}_j = \bm{y}_j$ the observed outcomes during the pre-experimental periods and $\bm{X}_j = \bm{x}_j$ the realizations of the covariates, we write the following quadratic program.
The decision variables to the quadratic program are $\bm{u}$ and $\bm{v}$, the weights of the synthetic control estimator.
\begin{align}
\min_{\substack{u_1, \ldots, u_n,\\ v_1, \ldots, v_n}}\quad & \bigg\|\overline{\bm{z}}-\sum_{j=1}^n u_j\bm{z}_j \bigg\|_2^2 + \bigg\|\overline{\bm{z}} - \sum_{j=1}^n v_j \bm{z}_j \bigg\|_2^2 \label{eqn:SynthControlExp} \\
\text{s.t. \ \ } 
& \sum_{j=1}^n u_j = 1,\nonumber \\
& \sum_{j=1}^n v_j = 1, \nonumber \\
& u_j, v_j \geq 0, \quad \forall j \in [n], \nonumber \\
& u_j v_j = 0, \quad \forall j \in [n], \nonumber \\
& \| u_j \|_0 \leq k. \nonumber
\end{align}
After solving \eqref{eqn:SynthControlExp}, we conduct an experiment and assign units that have a positive $u_j > 0$ weight into the treatment group and units that have a zero $u_j = 0$ weight into the control group.
Out of those units in the control group, we use the units that have a positive $v_j > 0$ weight to form the synthetic control unit.
We refer to the above design of experiment as the synthetic control design.
The synthetic control design as shown in \eqref{eqn:SynthControlExp}, as well as other formulations of the synthetic control design, such as the ones given in \citet{doudchenko2019designing, doudchenko2021synthetic}, is computationally challenging.
Recently, efforts have been made to computationally solve such problems \citep{lu2022synthetic}.

Denote the optimal solution to \eqref{eqn:SynthControlExp} as $(\bm{u}^*, \bm{v}^*)$.
The quality of the optimal solution $(\bm{u}^*, \bm{v}^*)$ will have an impact on the performance of the estimator $\widehat{\tau}^{SC}_t(\bm{u}^*, \bm{v}^*)$.
Before formally quantifying the performance of the estimator, we introduce a few more notations.
Denote $\overline{\beta} = \max_{j \in [n], t \in [T], w \in \{0,1\}} \beta_{jt}(w)$ and $\overline{\lambda} = \max_{j \in [n], t \in [T], w \in \{0,1\}} \lambda_{jt}(w)$.
Denote $\bblambda(0)$ to be a $(T_0 \times r)$ matrix whose $t$-th row is equal to $\bm{\lambda}_t(0)^\top$.
Denote $\zeta_{T_0}$ to be the smallest eigenvalue of $\bblambda(0)^\top \bblambda(0)$.

\begin{theorem}[\citet{abadie2021synthetic}]
\label{thm:SynthControlExp}
Assume that with probability one, the optimal solution $(\bm{u}^*, \bm{v}^*)$ is such that the covariates satisfy
\begin{subequations}
\begin{align*}
\Big\| \sum_{j=1}^n u^*_j \bm{X}_{j} - \sum_{j=1}^n f_j \bm{X}_{j} \Big\|_2^2 \leq d c^2, \qquad \Big\| \sum_{j=1}^n v^*_j \bm{X}_{j} - \sum_{j=1}^n f_j \bm{X}_{j} \Big\|_2^2 \leq d c^2,
\end{align*}
and the observed outcomes during the pre-experimental periods satisfy
\begin{align*}
\Big\|\sum_{j=1}^n u^*_j \bm{Y}_j - \sum_{j=1}^n f_j \bm{Y}_j \Big\|_2^2 \leq T_0 c^2, \qquad \Big\|\sum_{j=1}^n v^*_j \bm{Y}_j - \sum_{j=1}^n f_j \bm{Y}_j \Big\|_2^2 \leq T_0 c^2,
\end{align*}
\end{subequations}
where $c$ is a non-negative constant that does not depend on $T_0$.
In addition, assume $r \leq T_0$, and $\underline{\zeta} = \frac{\zeta_{T_0}}{T_0} > 0$ is a positive constant that does not depend on $T_0$.
Then,
\begin{align*}
\Big\vert \bE\big[ \widehat{\tau}^{SC}_t(\bm{u}^*, \bm{v}^*) - \tau_t \big] \Big\vert \leq 2 \bigg( \overline{\beta} d + \big(1 + \overline{\beta} d \big) \frac{\overline{\lambda}^2 r}{\underline{\zeta}} \bigg) c + \frac{2 \overline{\lambda}^2 r}{\underline{\zeta}} (2 \log{2n})^\frac{1}{2} \ \overline{\sigma} \ T_0^{-\frac{1}{2}}.
\end{align*}
\end{theorem}

Theorem~\ref{thm:SynthControlExp} suggests that, the ex-post bias of the synthetic control estimator $\widehat{\tau}^{SC}_t(\bm{u}^*, \bm{v}^*)$ consists of two parts.
The first part depends linearly on $c$, which reflects the quality of $(\bm{u}^*, \bm{v}^*)$ in minimizing the quadratic program \eqref{eqn:SynthControlExp}.
The second part depends on the random noises, and it decreases as $T_0$, the number of pre-experimental periods, increases.

To conclude this section, we note that Theorem~\ref{thm:SynthControlExp} only presents an upper bound of the ex-post bias of the synthetic control estimator; it does not fully characterize the true ex-post bias.
Yet it is probably the best characterization in the synthetic control literature.
Sometimes when the performance of the estimator is challenging to directly optimize, we optimize its proxies instead.

\section{Three Active Research Directions}
\label{sec:Survey}

The classical causal inference literature makes Assumptions~\ref{asp:NonInterference}~--~\ref{asp:RandomAssignment}.
So far, we have seen different optimization frameworks under these three classical assumptions.
These assumptions may not always hold in modern applications. 
Below we survey recent developments in experimental design for causal inference when these classical assumptions are violated.

\subsubsection*{Violation of Assumption~\ref{asp:NonInterference}}
Violation of Assumption~\ref{asp:NonInterference} leads to a rich literature on interference \citep{hudgens2008toward, tchetgen2012causal}. 
Intuitively, Assumption~\ref{asp:NonInterference} fails because the treatment assignment of one unit may have an impact on the potential outcomes of other units.
In the full generality, it requires all $n$ treatment assignments to describe the potential outcomes of each unit.

Depending on the causes, interference may be modeled in different ways.
Probably the most popular way of modeling interference is through a network, which leads to the network interference literature.
The network interference literature models each unit as a vertex on a network.
The treatment assignment on one unit may have an impact on the potential outcomes of other units through the edges of the network, which is a phenomenon called ``spillover'' effects.
One popular approach of studying network interference is through a concept called ``exposure mapping'' introduced by \citet{aronow2017estimating}, which develops the ``constant treatment response'' assumption introduced by \citet{manski2013identification}.
Exposure mapping is a dimension reduction mapping that reduces the dependence from all $n$ treatment assignments to a much smaller number of quantities.
See Example~\ref{exa:ExposureMapping} for an illustration of exposure mapping.

\begin{example}[Exposure Mapping]
\label{exa:ExposureMapping}
Let there be a graph $(V,E)$, where $V$ stands for the set of vertices and $E$ stands for the set of edges.
For any unit $j \in V$, let $\cN(j) = \big\{i \vert (i,j) \in E \big\}$ be the set of neighboring units.
We consider one exposure mapping that is the fraction of neighboring units who receive treatment (the ``distributional interactions'' assumption in \citet{manski2013identification}; see also \citet{athey2018exact, li2022random}).
For each unit $j \in V$, let the exposure mapping $d_j : \{0,1\}^{n-1} \to [0,1]$ be 
\begin{align*}
d_j(\bm{w}_{-j}) = \dfrac{\sum_{i \in \cN(j)} \bI\{w_i = 1\}}{\vert \cN(j) \vert},
\end{align*}
where $\bm{w}_{-j}$ stands for a vector of treatment assignments except for unit $j$.
Under such an exposure mapping, we can write out the potential outcomes by writing $Y_j(w_j, d_j(\bm{w}_{-j}))$ instead of writing $Y_j(\bm{w})$.
The potential outcomes become a two-dimensional function of the treatment assignment $w_j$ and the exposure mapping $d_j(\bm{w}_{-j})$.
\end{example}

Example~\ref{exa:ExposureMapping} only illustrates one specification of exposure mapping.
Exposure mapping can be specified in various other ways.
When the exposure mapping is well-specified and known, the network interference literature has examined different designs of experiments.
Most of these experiments share one similar idea and can be seen as some variants of ``cluster randomized experiments.''
In a cluster randomized experiment, all the units (vertices) are first grouped into multiple clusters.
Then, each cluster is randomly assigned into treatment or control, so that all the units within the same cluster receive the same version of treatment assignments.
See \citet{brennan2022cluster, candogan2023correlated, cortez2023exploiting, eckles2016estimating, eckles2017design, eichhorn2024low, jagadeesan2020designs, han2024population, harshaw2023design, holtz2020limiting, holtz2024reducing, jiang2023causal, 
leung2022rate, leung2023design, ni2023design, pouget2018optimizing, pouget2019variance, qu2021efficient, rolnick2019randomized, saveski2017detecting, ugander2013graph, ugander2023randomized, viviano2020experimental, viviano2023causal}, and references therein.

Sometimes the exposure mapping is unknown or potentially misspecified.
Recent works have examined various estimation strategies without using knowledge of the exposure mapping.
Yet, they need to assume that spillover effects decay with respect to the distance on the network; that is, spillover effects are only local.
See \citet{belloni2022neighborhood, leung2022causal, leung2022rate, savje2021average, yu2022estimating, yuan2021causal, yuan2023two}.
Because exposure mapping is unknown, the works in this literature usually conduct simple experiments and combine them with non-trivial analysis, with the exception of \citet{leung2022rate}, who studies how to choose the rate-optimal design of experiment.
If one can conduct multiple experiments on the same network over time, recent works consider rolling out experiments over time to make estimation and inference in the presence of unknown interference.
See \citet{boyarsky2023modeling, cortez2024combining, han2022detecting}.

Other than network interference, the interference literature has gained increasing popularity in modern marketplace applications, which leads to the literature on marketplace interference.
Marketplace interference can sometimes be modeled as network interference.
For example, the market equilibrium prices or the distributional interactions (Example~\ref{exa:ExposureMapping}) can sometimes be used as exposure mappings.
See \citet{bajari2021multiple, munro2021treatment, munro2024treatment, wager2021experimenting}.
On the other hand, marketplace interference can sometimes be captured by explicit models, such as auction models, discrete choice models, game-theoretic models, market-making models, and queuing models.
See \citet{basse2016randomization, bright2022reducing, dhaouadi2023price, johari2022experimental, kuang2024detecting, liao2023statistical, li2022interference, li2023experimenting, si2023optimal, ye2023cold}.

Recently, the interference literature also gains increasing popularity in recommendation systems, which leads to the literature on feedback loops, or symbiosis biases.
See \citet{goli2023bias, holtz2023study, johnson2017ghost, si2023tackling, zhan2024estimating}.
In a recommendation system, for example, user data obtained under previous recommendations are used as inputs to re-train the recommendation system.
Unless treatment and control data are separately used to train two systems, interference occurs from sharing a common pool of training data.
See \citet{si2023tackling} for an excellent review of feedback loops and related interference literature.

One remarkable special case of interference is the literature on temporal experiments, in which we can find temporal analogues of network interference and marketplace interference.
The temporal analogue of network interference is switchback experiments, which have appeared under various different names such as n-of-1 trials \citep{liang2023randomization}, time series experiments \citep{bojinov2019time}, and crossover designs \citep{basse2023minimax}.
A switchback experiment is a special case of network interference because the graph of interference in a switchback experiment follows a single line.
The spillover effects are referred to as ``carryover effects'' in this setting.
In earlier works, switchback experiments are used in agricultural applications to compare the effects of different feeding plans on milk yields \citep{cochran1941double}.
Recently, switchback experiments have gained increasing popularity as a result of the rise of modern applications such as on-demand service platforms (e.g., DoorDash, Lyft, and Uber; see \citet{chamandy2016experimentation, kastelman2018switchback} for blog posts on this topic).
\citet{glynn2020adaptive} is the first to study the design of switchback experiments in such modern applications, followed by \citet{bojinov2023design, chen2023switchback, hu2022switchback, jia2023faster, xiong2024data}.
See \citet{hu2022switchback} for an excellent introduction of switchback experiments.

The temporal analogue of marketplace interference involves modeling the carryover effects.
Two popular modeling approaches include Markov chain modeling and user modeling.
In Markov chain modeling, treatment assignments in the past will affect the outcomes in the future through intermediate states that evolve as a Markov chain \citep{farias2022markovian}.
If the treatment assignments in the past do not directly affect the outcomes in the future other than going through the intermediate states, such intermediate states are referred to as ``surrogates'' \citep{athey2019surrogate, prentice1989surrogate}.
Similar to the overlap of network interference with marketplace interference, Markovian interference overlaps with switchback experiments given the temporal nature of conducting experiments.
See \citet{glynn2020adaptive, hu2022switchback, jia2023faster}.
These aforementioned works focus on designing non-trivial experiments.
Other works focus on conducting simple random experiments but combining them with non-trivial analysis.
Some conduct experiments on one single unit \citep{liang2023randomization};
some on multiple units \citep{athey2019surrogate, farias2022markovian, huang2023estimating, wen2024analysis, yang2020targeting}.
In user modeling, the treatment effects are modeled using specific agent level models.
See \citet{hohnhold2015focusing, munro2023causal}.

\subsubsection*{Violation of Assumption~\ref{asp:iid}}

Violation of Assumption~\ref{asp:iid} usually concerns violating the identical distribution assumption, which leads to many directions in the literature such as generalizability.
This is an important topic in causal inference, yet we omit this topic from this manuscript.
Instead, we discuss the literature on treatment heterogeneity. 
This literature usually uses covariates to better describe treatment heterogeneity and uses Assumption~\ref{asp:iid-X} to overcome the violation of Assumption~\ref{asp:iid}.
As we have seen in Section~\ref{sec:Uncertainty2}, the value of covariates lies not only in reducing bias (through identifying all confounders) but also in reducing variance (through explaining variations coming from different covariates).

Using covariates in causal inference has a long history, probably as long as the causal inference literature itself.
The earlier books of \citet{cochran1948experimental, cox2000theory} have formally introduced classical designs of experiments, such as factorial design, block randomization, and stratified randomized design.
As \citet{rubin2008comment} commented, it is important to balance covariates in handling heterogeneity in randomized experiments.
Two of the most popular ways to balance covariates are through stratified randomized design (Definition~\ref{defn:Stratified} in Section~\ref{sec:Uncertainty2}) and through regression based methods (Section~\ref{sec:Uncertainty3}).

In line with stratified randomized design, there are numerous ways to partition units into strata.
See \citet{cytrynbaum2021designing, greevy2004optimal, higgins2016improving, lu2011optimal, tabord2023stratification}.
One special case when time is modeled as a covariate, \citet{deng2013improving, jin2023toward, tang2020control, wu2022non} study how to reduce variance in temporal experiments.
These above works focus on designing non-trivial experiments.
More generally, there are, of course, more extensive works that focus on conducting simple random experiments but combining them with non-trivial analysis such as matching, either directly on the covariates or on the propensity score or estimated propensity score.
We are unable to survey this rich literature and only refer to a few papers such as \citet{abadie2012martingale, abadie2006large, bai2022optimality, bertsimas2015power, dehejia2002propensity, diamond2013genetic, hirano2003efficient, imai2009essential, kallus2018optimal, rosenbaum1983central, rosenbaum1984reducing, rosenbaum1989optimal, zubizarreta2012using}, and references therein.

In line with regression based methods, Section~\ref{sec:Uncertainty3} is only a microcosm of the rich literature.
We have only examined the $D_A$-optimal and $D$-optimal criteria in Section~\ref{sec:Uncertainty3}.
There are many more optimal criteria in the literature (e.g., A-optimal and E-optimal criteria), which take different perspectives to scalarize the covariance matrix \eqref{eqn:ErrorCovarianceMatrix}.
We are unable to survey this rich literature and only refer to textbooks such as \citet{atkinson2007optimum, fedorov2013theory, pukelsheim2006optimal, silvey2013optimal}, and papers such as \citet{alexanderian2014optimal, atkinson1975optimal, card1993minimum, de2019approximate, ruan2021linear, titterington1975optimal, ucinski2005t}, and references therein.
Solving the optimization problems under different optimal criteria is computationally challenging.
One powerful tool to solve such optimization problems is semidefinite programming \citep{ahmadi2012convex, helmberg2002semidefinite, luo2010semidefinite, nie2014truncated, parrilo2003semidefinite, todd2001semidefinite, vandenberghe1996semidefinite}.
Recent works have also studied other objective functions that do not work with the covariance matrix \eqref{eqn:ErrorCovarianceMatrix}.
See \citet{chattopadhyay2022balanced, harshaw2024balancing, li2015value, morris1979finite}.
The optimal experimental design literature can also be extended to stepped wedge designs where a treatment is sequentially rolled out over a number of time periods.
See \citet{brown2006stepped, hemming2015stepped, hussey2007design, li2018optimal, xiong2023optimal}.
% A special case of stepped wedge design is the synthetic control design, where a small number of units receive treatment starting at the same time and over a number of time periods (Table~\ref{tbl:PanelData}).
% See \citet{abadie2021synthetic, doudchenko2019designing, doudchenko2021synthetic}.

Recently, as a result of the rise of clinical trial applications with sequential patient enrollment, covariate balancing problems are examined from the perspective of sequentially revealed covariate information.
The adaptive clinical trial literature refers to this perspective as ``covariate-adaptive'' experiments, as the treatment assignments of future units depend on the covariates and treatment assignments of past units.
The treatment assignments of future units do not depend on the observed outcomes of past units, so they are not ``response-adaptive'' \citep{hu2006theory}.
This literature stems from Efron's ``biased coin'' design \citep{efron1971forcing}.
Subsequently, \citet{pocock1975sequential} and \citet{atkinson1982optimum} develop general covariate-adaptive versions that make this literature popular.
For recent developments, see \citet{atkinson1982optimum, atkinson1999optimum, bertsimas2019covariate, bhat2020near, kapelner2014matching, rosenberger2008handling, rosenberger2015randomization, zhao2024pigeonhole}, and references therein.

\subsubsection*{Violation of Assumption~\ref{asp:RandomAssignment}}

Violation of Assumption~\ref{asp:RandomAssignment} leads to many directions in the literature.
One direction among them is the literature on response-adaptive experiments, or adaptive experiments for short \citep{hu2006theory}.
In an adaptive experiment, the units are sequentially enrolled into the experiment.
Following convention, unit $1$ arrives first, followed by unit $2$, and the last being unit $n$.
In the setting with no covariate, $W_j$, the treatment assignment for each unit $j \in [n]$, depends on $Y_1, ..., Y_{j-1}$ the observed outcomes of the past units as well as $W_1, ..., W_{j-1}$, the past treatment assignments.
Adaptive experiments are known to improve statistical efficiency and are thus desirable to experimenters \citep{murphy2005experimental, offer2021adaptive}.
However, adaptive experiments sometimes violate Assumption~\ref{asp:RandomAssignment}.
Consider the following two examples.

\begin{example}
\label{exa:AdaptiveExperiment1}
Consider a setting with $n=3$, and $Y_j(w), \forall j \in [n], w \in \{0,1\}$, the potential outcomes of all three units under both treatment and control are independent and identically sampled from the same $Bern(\frac{1}{2})$ Bernoulli distribution.
The experiment is conducted in two stages.
In stage one, we conduct a completely randomized experiment for units $1$ and $2$ such that exactly one of the two units receives treatment.
In stage two, unit $3$ receives treatment (or control) if treatment (or control) has a higher empirical performance.
In case of a tie, break the tie evenly.

Suppose we would like to estimate the expectation of the potential outcome under treatment, $\bE[Y(1)]$, by using the difference-in-means estimator (also referred to as the sample-mean estimator in the adaptive experiment literature), $\widehat{\mu}(1) = \frac{1}{N(1)} \sum_{j=1}^n Y_j \bI\{W_j=1\}$ where $N(1) = \sum_{j=1}^n \bI\{W_j=1\}$.
In this example, we can hand calculate all trajectories of sample paths.
We will then find $\bE\big[\widehat{\mu}(1)\big] = \frac{7}{16} < \frac{1}{2}$; that is, the difference-in-means estimator is biased.
This bias is caused by the adaptive sampling rule that favors treatment (or control) in stage two only if treatment (or control) has higher empirical performance in stage one.
\end{example}

\begin{example}
\label{exa:AdaptiveExperiment2}
Consider a survey sampling setting with an infinite number of units $1,2,...$, and only one version of treatment instead of both treatment and control.
Because there is only one version of treatment, there is only one version of potential outcome (which is observable) for each unit.
Let $Y_j, \forall j \in \{1,2,...\}$ the potential outcomes of all units be independent and identically sampled from the same $Bern(\frac{1}{2})$ Bernoulli distribution.
We conduct experiments one by one, and keep updating the difference-in-means estimator; that is, after we have conducted an experiment on unit $j$, we can calculate $\widehat{\mu}_{j} = \frac{1}{j} \sum_{i=1}^{j} Y_i$.

The experiment is adaptively conducted with the following stopping rule: stop at the first time when $\widehat{\mu}_{j} \geq \frac{2}{3}$.
We can define the stopping time $\tau$ as $\tau = \inf \big\{ j \big\vert \widehat{\mu}_{j} \geq \frac{2}{3} \big\}$.
Borrowing concepts from random walks, we can show that with positively probability, this sequence of experiments stops.
We can also easily see that, conditional on the sequence of experiments stops, the difference-in-means estimator has an expectation of at least $\bE\big[ \widehat{\mu}_{\tau} \big] \geq \frac{2}{3}$; that is, the difference-in-means estimator is biased.
This bias is caused by the adaptive stopping rule.
\end{example}

Examples~\ref{exa:AdaptiveExperiment1} and~\ref{exa:AdaptiveExperiment2} are only the two simplest examples.
More generally, because $W_j$, the treatment assignment for a unit, is correlated with $Y_1,...,Y_{j-1}$, the observed outcomes of the past units, it is also correlated with $(Y_1(1), Y_1(0)), ..., (Y_{j-1}(1), Y_{j-1}(0))$, the potential outcomes of the past units (unless some additional assumptions are made), thus violating Assumption~\ref{asp:RandomAssignment}.
Because Assumption~\ref{asp:RandomAssignment} is violated, using the standard difference-in-means estimator can be biased.
See \citet{bowden2017unbiased, deshpande2018accurate, dimakopoulou2021online, hadad2021confidence, hirano2023asymptotic, nie2018adaptively, shin2019sample, shin2019bias, zhan2021off, zhan2023policy, zhang2020inference, zhang2021statistical} under various settings and various adaptive policies.

The phenomenon shown in Example~\ref{exa:AdaptiveExperiment1} is referred to as ``optimistic sampling'' bias in \citet{shin2019sample}.
To correct for such a bias in estimation, two typical solutions are concerned with using suitable estimators and designing suitable experiments.
These two solutions can, of course, be combined.
We next introduce Assumption~\ref{asp:SequentialRandomAssignment}, the sequential random assignment assumption, which is usually satisfied under these two solutions.

\begin{assumption+}{\ref{asp:RandomAssignment}$^{**}$}[Sequential random assignment]
\label{asp:SequentialRandomAssignment}
For each unit $j \in [n]$, conditional on the history $\big\{ Y_i, W_i \big\}_{i=1}^{j-1}$, the treatment assignment $W_j$ and the pair of potential outcomes $(Y_j(1), Y_j(0))$ are independent, that is,
\begin{align*}
(Y_j(1), Y_j(0)) \independent W_j \ \vert \ \big\{ Y_i, W_i \big\}_{i=1}^{j-1}.
\end{align*}
\end{assumption+}

Starting from using suitable estimators, the IPW estimator (Definition~\ref{defn:IPW}) and the aggregate estimator (Definition~\ref{defn:Agg}) are two popular estimators to use.
As long as Assumptions~\ref{asp:SequentialRandomAssignment} (i.e., the sequential random assignment assumption) and Assumption~\ref{asp:Positivity} (i.e., the positivity assumption) hold, the IPW estimator is unbiased.
See \citet{bojinov2019time, zhan2023policy}, and references therein.
The aggregate estimator requires that the experiment is conducted in stages, such as in Example~\ref{exa:AdaptiveExperiment1}.
See \citet{zhang2020inference} and references therein.
Additionally, the simulations literature has studied another type of estimation strategy when the potential outcomes are modeled to have specific parametric forms.
See \citet{asmussen2007stochastic, glasserman2004monte, ross2013simulation} for classical textbooks and references therein.

Moving on to designing suitable experiments, we start with Assumption~\ref{asp:Positivity}.
Assumption~\ref{asp:Positivity} is required for the IPW estimator to be well defined.
As long as the IPW estimator is well defined, we can apply Assumption~\ref{asp:SequentialRandomAssignment} and martingale theorems to show that the IPW estimator is unbiased.
One critical idea to make the IPW estimator unbiased is to enforce Assumption~\ref{asp:Positivity}.
This can be done by enforcing an adaptive experiment to uniformly ``explore'' all the treatments with a strictly positive probability.
Not only does this idea work for the IPW type of estimators, it also proves to be effective for other types of estimators as well.
See \citet{bojinov2019time, bowden2017unbiased, dimakopoulou2021online, hadad2021confidence, ham2023designa, ham2023designb, zhan2021off, zhan2023policy, zhang2020inference, zhang2021statistical}.

On the other hand, one could argue that many of the adaptive policies in the above works are not for the purpose of estimation and inference, but rather for a purpose called ``(cumulative) regret minimization,'' which is a partial cause of biases.
We are unable to survey the rich literature on online learning, and only point to three alternative objectives called ``best-arm identification'' and ``simple regret minimization'' \citep{abbasi2018best, adusumilli2022minimax, audibert2010best, bubeck2009pure, chen2017adaptive, chen2018optimal, chen2023active, kasy2021adaptive, kato2022best, mannor2004sample, miao2023personalized, naby2024, qin2017improving, qin2022open, russo2016simple, tang2022offline, xu2023online, wu2022adaptive, zhang2024deep, zhou2014optimal}, as well as ``variance minimization'' \citep{antos2010active, armstrong2022asymptotic, aznag2024active, blackwell2022batch, carpentier2011finite, carpentier2011upper, carpentier2012minimax, carpentier2015adaptive, dai2024clip, deep2023asymptotically, fontaine2021online, grover2009active, hahn2011adaptive, russac2021b, wei2023adaptive, wei2024fair, xiong2023optimal, zhao2023adaptive}.
All these objectives are highly related (under certain assumptions some of them are even equivalent), and are often referred to as the pure-exploration objective.
When both the regret minimization objective and the pure-exploration objective are combined, there will be trade-offs \citep{athey2022contextual, bui2011committing, drugan2013designing, erraqabi2017trading, krishnamurthy2024proportional, qin2024optimizing, simchi2023multi, yao2021power, zhong2021achieving}.
See \citet{qin2024optimizing} for an excellent survey of the literature and a state-of-the-art framework without covariates.

The phenomenon shown in Example~\ref{exa:AdaptiveExperiment2} is referred to as ``stopping'' bias in \citet{shin2019sample}.
To correct for such a bias in estimation, we could borrow the same ideas from the two solutions in correcting for the optimistic sampling bias.
Motivated by improving statistical efficiency and reducing sample size, many works especially focus on studying how to stop the experiment as early as possible.
These works are collectively referred to as the sequential testing literature.
One classical technique that is popular in this literature is called the ``law of iterated logarithm.''
See \citet{siegmund2013sequential, wald2004sequential} for classical textbooks, and \citet{bibaut2022near, cho2024peeking, jamieson2018bandit, johari2015always, johari2017peeking, lindon2022anytime, liang2023experimental, malenica2023anytime, ramdas2020admissible, ramdas2023game} for recent developments under various settings.

\section{Conclusion}
\label{sec:Conclusion}

This manuscript focuses on experimental design problems that arise in causal inference contexts as viewed through an optimization lens.
We discuss three major frameworks of experimental design problems in details: the robust optimization framework, the stochastic optimization framework, and the deterministic optimization framework. 
The first framework models the uncertainty of potential outcomes to be more ambiguous, wherein the optimal design is usually random.
Conversely, the second framework postulates distributional knowledge and the third framework postulates well-specified models to capture the uncertainty of potential outcomes, wherein the optimal design is usually deterministic. 

These different frameworks reflect different ways to model the uncertainty governing the potential outcomes.
For an experimenter to choose from one of the three frameworks, we recommend choosing the appropriate framework based on the uncertainty that the experimenter faces.
We distinguish three cases.
In the first case when we have strong knowledge to explain the uncertainty, it is appropriate to adopt the deterministic optimization framework.
As an example given in \citet{rubin1978bayesian}, this could be the case when an industrial experiment comparing manufacturing procedures may have strong knowledge about the relationship between covariates and potential outcomes.
In the second case when we have some knowledge to describe a prior distribution of the potential outcomes, it is appropriate to adopt the stochastic optimization framework.
This could be the case when we have collected historical data to estimate the prior distribution, and when we believe Assumption~\ref{asp:iid} to hold; that is, there is no distributional shift.
Finally, in the third case when we have little knowledge about the potential outcomes, it is appropriate to adopt the robust optimization framework.
This could be the case when we have no historical data, or, as commented in \citet{wu1981robustness}, because ``the experimenter's knowledge about the (potential outcomes) model is never perfect.''

Depending on how we model uncertainty, it is conceptually simple to cast experimental design problems as optimization problems under one of the three frameworks.
This manuscript presents, in consequence, a range of potential research opportunities to study experimental design problems for causal inference through an optimization lens.

% \section*{Acknowledgments}
% \label{sec:Acknowledgment}

% The author would like to thank Lihua Lei, who has shared valuable perspectives that built the foundation of this manuscript. 
% The author would like to thank Alberto Abadie, David Alderson, Volodymyr Babich, Yuchen Hu, Stefanus Jasin, Ramesh Johari, Shuangning Li, Tu Ni, Chao Qin, Christopher Ryan, Nian Si, Johan Ugander, Ruoxuan Xiong, Ruohan Zhan, Bo Zhang, course participants at the National University of Singapore, and five anonymous referees, whose suggestions and feedbacks have significantly improved this manuscript. 
% This manuscript builds on the author's lecture notes taught at the National University of Singapore.

% Appendix optional
%\APPENDIX{This is the Appendix Title} 
%\APPENDIX{}  % <-- This is an appendix without a title.

%\section{References}
% Make bibliography with BibTeX
% (see bibexample.tex at TutORials web site)
% Use TutORials.bst
% Note that there are no journal and proceedings abbreviations;
%   other requirements given in bibexample

\setcitestyle{numbers} % set the citation style to ``numbers''.
\bibliographystyle{informs2014}  % put ./TutORials.bst if using locally
\bibliography{bibliography}    % put your bib file name here

\clearpage
\begin{APPENDIX}{Proofs}
\renewcommand\thefigure{\thesection\arabic{figure}}    
\renewcommand\thetable{\thesection\arabic{table}}    
\setcounter{figure}{0}    
\setcounter{table}{0}

\section{Proofs from Section~\ref{sec:ExpDesign}}

\proof{Proof of Lemma~\ref{lem:Unbiasedness}.}
Let $\bm{0}$ and $\bm{1}$ be two vectors of zeros and ones, respectively.
To start, we prove the first claim $\bE_{\eta}\big[ \widehat{\tau}^{DM} \vert N(1) \notin \{0,n\} \big] = \tau_{\bm{Y}(1), \bm{Y}(0)}$ under the Bernoulli design.
We start by focusing on the first component of $\widehat{\tau}^{DM}$,
\begin{align*}
\bE_\eta\bigg[ \frac{\sum_{j=1}^n Y_j \bI\{W_j=1\}}{\sum_{j=1}^n \bI\{W_j=1\}} & \Big\vert N(1) \notin \{0,n\} \bigg] \\
= & \ \sum_{\substack{\bm{w}\in\{0,1\}^n \\ \bm{w}\ne\bm{0},\bm{1}}} \frac{\prod_{j=1}^n p^{\bI\{w_j=1\}} (1-p)^{\bI\{w_j=0\}}}{1- p^n - (1-p)^n}  \cdot \frac{\sum_{j=1}^n Y_j(1) \bI\{w_j=1\}}{\sum_{j=1}^n \bI\{w_j=1\}} \\
= & \ \frac{1}{1- p^n - (1-p)^n} \sum_{j=1}^n Y_j(1) \cdot \sum_{m=1}^{n-1} \frac{1}{m} \ p^m (1-p)^{n-m} \dbinom{n-1}{m-1} \\
= & \ \frac{1}{1- p^n - (1-p)^n} \sum_{j=1}^n Y_j(1) \cdot \frac{1}{n} \sum_{m=1}^{n-1} p^m (1-p)^{n-m} \dbinom{n}{m} \\
= & \ \frac{1}{1- p^n - (1-p)^n} \sum_{j=1}^n Y_j(1) \cdot \frac{1}{n} \cdot \big( (p + (1-p))^n - p^n - (1-p)^n \big) \\
= & \ \frac{1}{n} \sum_{j=1}^n Y_j(1).
\end{align*}
The first equality is because conditional on $N(1) \notin \{0,n\}$, we need to normalize the probability mass functions by $1-p^n-(1-p)^n$.
The second equality is using a different way to count the summation, by counting how many occurrences of $Y_j(1)$ are there when denominator is equal to $n$.
Similar to the above derivation, the second component of $\widehat{\tau}^{DM}$ can be derived as 
\begin{align*}
\bE_\eta\Bigg[ \frac{\sum_{j=1}^n Y_j \bI\{W_j=1\}}{\sum_{j=1}^n \bI\{W_j=1\}} \Big\vert N(1) \notin \{0,n\} \Bigg] = \frac{1}{n} \sum_{j=1}^n Y_j(0).
\end{align*}
Combining two components we prove the first claim that $\bE_{\eta}\big[ \widehat{\tau}^{DM} \vert N(1) \notin \{0,n\} \big] = \tau_{\bm{Y}(1), \bm{Y}(0)}$.

The other three claims follow easily by definition.
Under the Bernoulli design,
\begin{align*}
\bE_{\eta}\big[ \widehat{\tau}^{IPW} \big] = & \ \frac{1}{n} \sum_{j=1}^n \frac{Y_j(1) \ p}{p} - \frac{1}{n} \sum_{j=1}^n \frac{Y_j(0) \ (1-p)}{1-p} \\
= & \ \frac{1}{n} \sum_{j=1}^n Y_j(1) - \frac{1}{n} \sum_{j=1}^n Y_j(0).
\end{align*}
The first equality is because of linearity of expectation.

Under the completely randomized design,
\begin{align*}
\bE_{\eta}\big[ \widehat{\tau}^{DM} \big] = & \ \frac{\sum_{j=1}^n Y_j(1) \ p}{pn} - \frac{\sum_{j=1}^n Y_j(0) \ (1-p)}{(1-p)n} \\
= & \ \frac{1}{n} \sum_{j=1}^n Y_j(1) - \frac{1}{n} \sum_{j=1}^n Y_j(0).
\end{align*}
The first equality is because the completely randomized design holds the denominators as fixed.

Under the completely randomized design,
\begin{align*}
\bE_{\eta}\big[ \widehat{\tau}^{IPW} \big] = & \ \frac{1}{n} \sum_{j=1}^n \frac{Y_j(1) \ p}{p} - \frac{1}{n} \sum_{j=1}^n \frac{Y_j(0) \ (1-p)}{1-p} \\
= & \ \frac{1}{n} \sum_{j=1}^n Y_j(1) - \frac{1}{n} \sum_{j=1}^n Y_j(0).
\end{align*}
Again the first equality is because of linearity of expectation.
\hfill \halmos
\endproof

\section{Proofs from Section~\ref{sec:Uncertainty1}}

\proof{Proof of Lemma~\ref{lem:CRDOptPermuInv}.}
Because of Assumption~\ref{asp:PermutationInvariance}, we know that $L(\bm{w}, \bm{g}) = L(\pi(\bm{w}), \pi(\bm{g}))$.
Then we have
\begin{align*}
\max_{\bm{g} \in \cG} r(\tilde{\eta}, \bm{g}) = & \ \max_{\bm{g} \in \cG} \sum_{\bm{w} \in \{0,1\}^n} \tilde{\eta}(\bm{w}) L(\bm{w}, \bm{g}) \\
= & \ \frac{1}{n!} \max_{\bm{g} \in \cG} \ \sum_{\pi \in \Pi} \ \sum_{\bm{w} \in \{0,1\}^n} \eta(\pi(\bm{w})) L(\bm{w}, \bm{g}) \\
\leq & \ \frac{1}{n!} \ \sum_{\pi \in \Pi} \ \max_{\bm{g} \in \cG} \sum_{\bm{w} \in \{0,1\}^n} \eta(\pi(\bm{w})) L(\bm{w}, \bm{g}) \\
= & \ \frac{1}{n!} \ \sum_{\pi \in \Pi} \ \max_{\bm{g} \in \cG} \sum_{\bm{w} \in \{0,1\}^n} \eta(\pi(\bm{w})) L(\pi(\bm{w}), \pi(\bm{g})) \\
= & \ \frac{1}{n!} \ \sum_{\pi \in \Pi} \ \max_{\bm{g} \in \cG} \ r(\eta_\pi, \bm{g}) \\
= & \ r(\eta, \bm{g}),
\end{align*}
where the second equality is because we expand $\tilde{\eta}$ and exchange the double summation;
the first inequality is because we exchange maximization and summation, so for each $\pi\in\Pi$ there might exist a different maximizing $\bm{g}$.
\hfill \halmos
\endproof

\proof{Proof of Theorem~\ref{thm:CRDOptPermuInv}.}
We first simplify the loss function.
For any $\bm{w}$, let $n(1), n(0)$ denote the number of units in the treatment and control groups, respectively.
\begin{align}
\bE_{\bm{\epsilon}} \big[(\widehat{\tau} - \tau)^2\big] = & \ \bE_{\bm{\epsilon}} \Bigg[\bigg(\frac{1}{n(1)} \sum_{j=1}^n \bI\{w_j=1\}(g_j+\epsilon_{j1}) - \frac{1}{n(0)} \sum_{j=1}^n \bI\{w_j=0\}(g_j+\epsilon_{j0})\bigg)^2\Bigg] \nonumber \\
= & \ \bigg(\frac{1}{n(1)} \sum_{j=1}^n \bI\{w_j=1\}g_j - \frac{1}{n(0)} \sum_{j=1}^n \bI\{w_j=0\}g_j\bigg)^2 + \nonumber \\
& \qquad \qquad \qquad \bE_{\bm{\epsilon}} \Bigg[\bigg(\frac{1}{n(1)} \sum_{j=1}^n \bI\{w_j=1\}\epsilon_{j1} - \frac{1}{n(0)} \sum_{j=1}^n \bI\{w_j=0\}\epsilon_{j0}\bigg)^2\Bigg] \nonumber \\
= & \ \bigg(\frac{1}{n(1)} \sum_{j=1}^n \bI\{w_j=1\}g_j - \frac{1}{n(0)} \sum_{j=1}^n \bI\{w_j=0\}g_j\bigg)^2 + \sigma^2 \bigg(\frac{1}{n(1)} + \frac{1}{n(0)}\bigg), \label{eqn:Wu1981Intermediate}
\end{align}
where the first equality is plugging in \eqref{eqn:Wu1981Model} and canceling $\alpha_1, \alpha_0$;
the second equality is expanding the square, and noticing that the cross term is equal to zero because the noises $\epsilon_{jw}, \forall j\in[n], w\in\{0,1\}$ have zero means;
the third equality is because of the noises $\epsilon_{jw}$ being independent across $j\in[n]$.

Then, because of Lemma~\ref{lem:CRDOptPermuInv}, we focus only on a distribution of completely randomized designs.
Define $\cW(n(1)) = \big\{\bm{w} \in \{0,1\}^n \big\vert \sum_{j=1}^n \bI\{w_j=1\} = n(1) \big\}$.
Apparently we do not wish to have no treatment or control units, so $n(1) \notin \{0, n\}$.
Next we have
\begin{align*}
r(\tilde{\eta}, \bm{g}) = & \ \sum_{\bm{w} \in \{0,1\}^n} \eta(\bm{w}) L(\bm{w}, \bm{g}) \\
= & \ \sum_{n(1) = 1}^{n-1} \Pr(\bm{W} \in \cW(n(1))) \sum_{\bm{w} \in \cW(n(1))} \delta(\bm{w}) L(\bm{w}, \bm{g}).
\end{align*}
Here $\sum_{\bm{w} \in \cW(n(1))} \delta(\bm{w})$ stands for the discrete probability distribution induced by the completely randomized design; that is, out of a total of $n$ many units, we randomly assign $n(1)$ of them in the treatment group.
Under this discrete probability distribution, we can calculate
\begin{align*}
\sum_{\bm{w} \in \cW(n(1))} \delta(\bm{w}) \bigg( \sum_{j=1}^n \bI\{w_j=1\}g_j \bigg)^2 & = \frac{n(1)}{n} \sum_{j=1}^n g_j^2 + \frac{n(1)(n(1)-1)}{n(n-1)} \sum_{i \ne j} g_i g_j, \\
\sum_{\bm{w} \in \cW(n(1))} \delta(\bm{w}) \bigg( \sum_{j=1}^n \bI\{w_j=0\}g_j \bigg)^2 & = \frac{n(0)}{n} \sum_{j=1}^n g_j^2 + \frac{n(0)(n(0)-1)}{n(n-1)} \sum_{i \ne j} g_i g_j, \\
\sum_{\bm{w} \in \cW(n(1))} \delta(\bm{w}) \bigg( \sum_{j=1}^n \bI\{w_j=1\}g_j \bigg) \cdot \bigg( \sum_{j=1}^n \bI\{w_j=0\}g_j \bigg) & = \frac{n(1) n(0)}{n(n-1)} \sum_{i \ne j} g_i g_j.
\end{align*}

Using the above expressions and using \eqref{eqn:Wu1981Intermediate}, we re-write the risk function to be
\begin{align*}
\sum_{\bm{w} \in \cW(n(1))} \delta(\bm{w}) L(\bm{w}, \bm{g}) = \bigg(\frac{1}{n(1)} + \frac{1}{n(0)}\bigg) \cdot \bigg( \frac{1}{n}\sum_{j=1}^ng_j^2 - \frac{1}{n(n-1)} \sum_{i\ne j} g_i g_j + \sigma^2\bigg).
\end{align*}
Because the second component of the risk function does not depend on $n(1), n(0)$, how we choose the sample sizes do not affect the second component.
To minimize the first component, we choose $n(1) = n(0) = \frac{n}{2}$.
\hfill \halmos
\endproof

\proof{Proof of Lemma~\ref{lem:UniBSetReduction}.}
Note that, we can re-write the risk function as
\begin{align*}
\bE_{\eta_{\bm{p}}} & \big[(\widehat{\tau} - \tau)^2\big] \\
= & \ \bE_{\eta_{\bm{p}}}\Bigg[\bigg(\frac{\sum_{j=1}^n y_j(1) \bI\{W_j=1\}}{n \Pr\{W_j=1\}} - \frac{\sum_{j=1}^n y_j(0) \bI\{W_j=0\}}{n \Pr\{W_j=0\}} - \frac{1}{n} \sum_{j=1}^n y_j(1) + \frac{1}{n} \sum_{j=1}^n y_j(0) \bigg)^2\Bigg] \\
= & \ \frac{1}{n^2} \bE_{\eta_{\bm{p}}}\Bigg[\bigg(\sum_{j=1}^n y_j(1) \Big(\frac{\bI\{W_j=1\}}{\Pr(W_j=1)} - 1\Big) - \sum_{j=1}^n y_j(0) \Big(\frac{\bI\{W_j=0\}}{\Pr(W_j=0)} - 1\Big) \bigg)^2\Bigg].
\end{align*}

We next expand the square.
\begin{align*}
\bE_{\eta_{\bm{p}}} & \big[(\widehat{\tau} - \tau)^2\big] \\
= & \ \frac{1}{n^2} \sum_{j=1}^n \Bigg\{ y_j(1)^2 \bE_{\eta_{\bm{p}}}\bigg[ \Big( \frac{\bI\{W_j=1\}}{\Pr(W_j=1)} -1 \Big)^2 \bigg] + y_j(0)^2 \bE_{\eta_{\bm{p}}}\bigg[ \Big( \frac{\bI\{W_j=0\}}{\Pr(W_j=0)} -1 \Big)^2 \bigg] \\
& \qquad \qquad \qquad \qquad \qquad \qquad - 2 y_j(1)y_j(0) \bE_{\eta_{\bm{p}}}\bigg[ \Big( \frac{\bI\{W_j=1\}}{\Pr(W_j=1)} -1 \Big) \Big( \frac{\bI\{W_j=0\}}{\Pr(W_j=0)} -1 \Big) \bigg] \Bigg\} \\
= & \ \frac{1}{n^2} \sum_{j=1}^n \Bigg\{ y_j(1)^2 \Big(\frac{1-p_j}{p_j}\Big) + y_j(0)^2 \Big(\frac{p_j}{1-p_j}\Big) + 2 y_j(1) y_j(0) \Bigg\} \\
= & \ \frac{1}{n^2} \cdot \sum_{j=1}^n \frac{1}{p_j(1-p_j)} \cdot \Big( y_j(1) (1-p_j) + y_j(0) p_j \Big)^2.
\end{align*}
Here the first equality is because, under the Bernoulli design, the treatment assignments of two units are independent.
Due to independence, for any $i \ne j \in [n]$ and any $w_i, w_j \in \{0,1\}$, we have 
\begin{multline*}
\bE_{\eta_{\bm{p}}}\bigg[ \Big( \frac{\bI\{W_i=w_i\}}{\Pr(W_i=w_i)} -1 \Big) \Big( \frac{\bI\{W_j=w_j\}}{\Pr(W_j=w_j)} -1 \Big) \bigg] \\
= \bE_{\eta_{\bm{p}}}\bigg[ \Big( \frac{\bI\{W_i=w_i\}}{\Pr(W_i=w_i)} -1 \Big) \bigg] \bE_{\eta_{\bm{p}}}\bigg[\Big( \frac{\bI\{W_j=w_j\}}{\Pr(W_j=w_j)} -1 \Big) \bigg] = 0.
\end{multline*}
So the cross-terms between any two units $i \ne j \in [n]$ are all equal to zero.
\hfill \halmos
\endproof

\proof{Proof of Theorem~\ref{thm:BDOptUniBSet}.}
The proof of Theorem~\ref{thm:BDOptUniBSet} is quite straightforward.
We first characterize the worst-case potential outcomes.
For any vector of treatment probabilities $\bm{p}$, we have for any $j \in [n]$,
\begin{align*}
\Big( y_j(1) (1-p_j) + y_j(0) p_j \Big)^2 \leq b^2,
\end{align*}
where the inequality takes equality when the potential outcomes $(y_j(1), y_j(0))$ are such that either $y_j(1) = y_j(0) = b$ or $y_j(1) = y_j(0) = -b$.

Under such worst-case potential outcomes, we can re-write the risk function as
\begin{align*}
\bE_{\eta_{\bm{p}}} \big[(\widehat{\tau} - \tau)^2\big] = \ \frac{b^2}{n^2} \cdot \sum_{j=1}^n \frac{1}{p_j(1-p_j)} \geq \ \frac{4 b^2}{n},
\end{align*}
where inequality takes equality when $p_j = \frac{1}{2}$ across all $j\in[n]$.
\hfill \halmos
\endproof

\section{Proofs from Section~\ref{sec:Uncertainty2}}

\proof{Proof of Lemma~\ref{lem:MatchedPairReduction}}
We introduce a few useful notations.
We first define the ex-ante bias conditional on $\bb{X} = \bb{x}$ as
\begin{align*}
\Bias^{ante}_{\cP}\big(\widehat{\tau} \vert \bb{x}\big) = \bE_{\eta_{\cP}, \cF^n \vert \bb{X}} \big[ \widehat{\tau} \vert \bb{x} \big] - \tau_{\bb{x}},
\end{align*}
and then the ex-post bias conditional on $\bb{X} = \bb{x}$ and $\bm{W} = \bm{w}$ as
\begin{align*}
\Bias^{post}_{\cP}\big(\widehat{\tau} \vert \bb{x}, \bm{w}\big) = \bE_{\cF^n \vert \bb{X}} \big[ \widehat{\tau} \vert \bb{x}, \bm{w} \big] - \tau_{\bb{x}}.
\end{align*}

It is immediate to make the following two observations.
First, the expectation of the ex-post bias over the randomized design $\eta_\cP$ is equal to the ex-ante bias, that is,
\begin{align}
\bE_{\bm{W} \sim \eta_\cP} \Big[ \Bias^{post}_{\cP}\big(\widehat{\tau} \vert \bb{x}, \bm{W}\big) \Big] = \Bias^{ante}_{\cP}\big(\widehat{\tau} \vert \bb{x}\big). \label{eqn:MatchedPair1}
\end{align}
Second, the ex-ante bias
\begin{align}
\Bias^{ante}_{\cP}\big(\widehat{\tau} \vert \bb{x}\big) = \frac{2}{n} \sum_{j=1}^n \Big( \bE_{\cF \vert \bm{X}_j}[Y_j(1) \vert \bm{x}_j] - \bE_{\cF \vert \bm{X}_j}[Y_j(0) \vert \bm{x}_j] \Big) - \tau_{\bb{x}} = 0, \label{eqn:MatchedPair2}
\end{align}
is zero (thus identical across different choices of $\cP$), because we conduct a completely randomized experiment within each stratum.

Similar to Lemma~\ref{lem:BiasVariance}, we can decompose the mean squared error in the objective function of \eqref{eqn:MatchedPairObj} as follows,
\begin{multline}
\bE_{\eta_\cP, \cF^n \vert \bb{X}} \big[ (\widehat{\tau} - \tau_{\bb{x}})^2 \vert \bb{x} \big] = \ \Big[ \Bias^{ante}_{\cP}\big(\widehat{\tau} \vert \bb{x}\big) \Big]^2 + \Var_{\eta_\cP, \cF^n \vert \bb{X}}\big( \widehat{\tau} \vert \bb{x} \big) \\
= \ \Big[ \Bias^{ante}_{\cP}\big(\widehat{\tau} \vert \bb{x}\big) \Big]^2 + \bE_{\eta_\cP}\Big[ \Var_{\cF^n \vert \bb{X}}\big(\widehat{\tau} \vert \bb{x}, \bm{W}\big) \Big] + \Var_{\eta_\cP}\Big( \bE_{\cF^n \vert \bb{X}}\big[\widehat{\tau} \vert \bb{x}, \bm{W}\big] \Big), \label{eqn:MatchedPairIntermediate}
\end{multline}
where the first equality is applying Lemma~\ref{lem:BiasVariance} and the second inequality is the law of total variance.
The third term in the above expression can be expressed as
\begin{align}
& \ \Var_{\eta_{\cP}}\Big( \bE_{\cF^n \vert \bb{X}}\big[\widehat{\tau} \vert \bb{x}, \bm{W}\big] \Big) \nonumber \\
= & \ \bE_{\eta_{\cP}} \Big[ \big( \bE_{\cF^n \vert \bb{X}}[\widehat{\tau} \vert \bb{x}, \bm{W}] - \bE_{\eta_{\cP}, \cF^n \vert \bb{X}}[\widehat{\tau} \vert \bb{x}, \bm{W}] \big)^2 \Big] \nonumber \\
= & \ \bE_{\eta_{\cP}} \Big[ \big( \Bias^{post}_{\cP}\big(\widehat{\tau} \vert \bb{x}, \bm{W}\big) - \Bias^{ante}_{\cP}\big(\widehat{\tau} \vert \bb{x}\big) \big)^2 \Big] \nonumber \\
= & \ \bE_{\eta_{\cP}} \Big[ \big( \Bias^{post}_{\cP}\big(\widehat{\tau} \vert \bb{x}, \bm{W}\big) \big)^2 \Big] - 2 \bE_{\eta_{\cP}} \Big[ \Bias^{post}_{\cP}\big(\widehat{\tau} \vert \bb{x}, \bm{W}\big) \Big] \cdot \Bias^{ante}_{\cP}\big(\widehat{\tau} \vert \bb{x}\big) + \big( \Bias^{ante}_{\cP}\big(\widehat{\tau} \vert \bb{x}\big) \big)^2 \nonumber \\
= & \ \bE_{\eta_{\cP}} \Big[ \big( \Bias^{post}_{\cP}\big(\widehat{\tau} \vert \bb{x}, \bm{W}\big) \big)^2 \Big] - \big( \Bias^{ante}_{\cP}\big(\widehat{\tau} \vert \bb{x}\big) \big)^2 \label{eqn:VarDecompose:PostAnteBiases}
\end{align}
where the first equality is the definition of variance;
the third equality is because conditional on the realization $\bb{X} = \bb{x}$, $\Bias^{ante}_{\cP}\big(\widehat{\tau} \vert \bb{x}\big)$ is constant;
the fourth equality is using \eqref{eqn:MatchedPair1}.

Putting \eqref{eqn:VarDecompose:PostAnteBiases} into \eqref{eqn:MatchedPairIntermediate} we have
\begin{align}
\bE_{\eta_\cP, \cF^n \vert \bb{X}} \big[ (\widehat{\tau} - \tau_{\bb{x}})^2 \vert \bb{x} \big] = \bE_{\eta_\cP}\Big[ \Var_{\cF^n \vert \bb{X}}\big(\widehat{\tau} \vert \bb{x}, \bm{W}\big) \Big] + \bE_{\eta_{\cP}} \Big[ \big( \Bias^{post}_{\cP}\big(\widehat{\tau} \vert \bb{x}, \bm{W}\big) \big)^2 \Big]. \label{eqn:MatchedPairFinal}
\end{align}

The first term from \eqref{eqn:MatchedPairFinal} is equal to 
\begin{align*}
& \ \bE_{\eta_\cP}\Big[ \Var_{\cF^n \vert \bb{X}}\big(\widehat{\tau} \vert \bb{x}, \bm{W}\big) \Big] \\
= & \ \bE_{\eta_\cP}\bigg[ \frac{4}{n^2} \sum_{j=1}^n \Big( \Var(Y_j(1) \vert \bm{x}_j) \bI\{W_j=1\} + \Var(Y_j(0) \vert \bm{x}_j) \bI\{W_j=0\} \Big) \bigg] \\
= & \ \frac{2}{n^2} \sum_{j=1}^n \Big( \Var(Y_j(1) \vert \bm{x}_j) + \Var(Y_j(0) \vert \bm{x}_j) \Big),
\end{align*}
which is identical across different choices of $\cP$.
The first equality holds because the variance of the sum of independent variables is equal to the sum of their respective variances.

Then from \eqref{eqn:MatchedPairFinal}, because its first term is identical across different $\cP$, we can show that the minimization problem as defined in \eqref{eqn:MatchedPairObj} is equivalent to
\begin{align*}
\min_{\cP} \ \bE_{\eta_{\cP}} \Big[ \big( \Bias^{post}_{\cP}\big(\widehat{\tau} \vert \bb{x}, \bm{W}\big) \big)^2 \Big].
\end{align*}

We then write
\begin{align*}
\bE_{\eta_{\cP}} \Big[ \big( \Bias^{post}_{\cP}\big(\widehat{\tau} \vert \bb{x}, \bm{W}\big) \big)^2 \Big] = & \ \Var_{\eta_{\cP}} \Big( \bE\big[ \widehat{\tau} \vert \bb{x}, \bm{W} \big] \Big) \\
= & \ \Var_{\eta_{\cP}} \Big( \frac{2}{n} \sum_{j=1}^n \Big( \bE[Y_j(1)\vert \bm{x}_j] \bI\{W_j=1\} - \bE[Y_j(0)\vert \bm{x}_j] \bI\{W_j=0\} \Big) \Big) \\
= & \ \frac{4}{n^2} \Var_{\eta_{\cP}} \Big( \sum_{j=1}^n \Big( \bE[Y_j(1)+Y_j(0)\vert \bm{x}_j] \bI\{W_j=1\} - \bE[Y_j(0)\vert \bm{x}_j] \Big) \Big) \\
= & \ \frac{4}{n^2} \Var_{\eta_{\cP}} \Big( \sum_{j=1}^n \Big( \bE[Y_j(1)+Y_j(0)\vert \bm{x}_j] \bI\{W_j=1\} \Big) \Big),
\end{align*}
where the first equality is using \eqref{eqn:MatchedPair1} and \eqref{eqn:MatchedPair2};
the second equality is using the definition of the difference-in-means estimator and that we conduct a completely randomized design within each stratum;
the last equality is because adding a constant does not change the variance.
\hfill \halmos
\endproof

\proof{Proof of Theorem~\ref{thm:MatchedPairOpt}.}
To prove Theorem~\ref{thm:MatchedPairOpt}, we will repeatedly refer to the following inequality.
For any four numbers $g_{j_1} \geq g_{j_2} \geq g_{j_3} \geq g_{j_4}$, we have
\begin{align}
(g_{j_1} - g_{j_4})^2 + (g_{j_2} - g_{j_3})^2 \geq (g_{j_1} - g_{j_2})^2 + (g_{j_3} - g_{j_4})^2. \label{eqn:MatchedPairUsefulIneq}
\end{align}
This inequality holds because it is equivalent to 
\begin{align*}
g_{j_1}g_{j_2} + g_{j_3}g_{j_4} \geq g_{j_1}g_{j_4} + g_{j_2}g_{j_3},
\end{align*}
which is then equivalent to
\begin{align*}
(g_{j_1} - g_{j_3})(g_{j_2} - g_{j_4}) \geq 0,
\end{align*}
which finishes the proof of inequality \eqref{eqn:MatchedPairUsefulIneq}.

Now we begin to prove Theorem~\ref{thm:MatchedPairOpt}.
It is useful to observe that for any $j \in [n]$,
\begin{align*}
\Var_{\eta_\cP}(\bI\{W_j=1\}) = \bE_{\eta_\cP}\big[\bI\{W_j=1\}^2\big] - \bE_{\eta_\cP}[\bI\{W_j=1\}]^2 = \frac{1}{4}.
\end{align*}

Then we move on to the covariance terms.
For any two units $i, j \in [n]$ but do not belong to the same stratum, because the randomization between different strata are independent, the covariance term $\Cov_{\eta_\cP}(\bI\{W_i=1\},\bI\{W_j=1\}) = 0$ is equal to zero.

\begin{table}[h]
\renewcommand{\arraystretch}{2}
\begin{tabular}{|c|c|c|c|c|c|}
\hline
$\frac{1}{4}$   & $-\frac{1}{20}$ & $-\frac{1}{20}$ & $-\frac{1}{20}$ & $-\frac{1}{20}$ & $-\frac{1}{20}$ \\ \hline
$-\frac{1}{20}$ & $\frac{1}{4}$   & $-\frac{1}{20}$ & $-\frac{1}{20}$ & $-\frac{1}{20}$ & $-\frac{1}{20}$ \\ \hline
$-\frac{1}{20}$ & $-\frac{1}{20}$ & $\frac{1}{4}$   & $-\frac{1}{20}$ & $-\frac{1}{20}$ & $-\frac{1}{20}$ \\ \hline
$-\frac{1}{20}$ & $-\frac{1}{20}$ & $-\frac{1}{20}$ & $\frac{1}{4}$   & $-\frac{1}{20}$ & $-\frac{1}{20}$ \\ \hline
$-\frac{1}{20}$ & $-\frac{1}{20}$ & $-\frac{1}{20}$ & $-\frac{1}{20}$ & $\frac{1}{4}$   & $-\frac{1}{20}$ \\ \hline
$-\frac{1}{20}$ & $-\frac{1}{20}$ & $-\frac{1}{20}$ & $-\frac{1}{20}$ & $-\frac{1}{20}$ & $\frac{1}{4}$   \\ \hline
\end{tabular}
\caption{An illustration of a block of the covariance matrix, when there are $6$ units within this stratum.}
\label{tbl:VarCovMatrix6}
\end{table}

For any two units $i, j \in [S_l] \subseteq [n]$ that belong to the same stratum, we can calculate
\begin{align*}
\Cov_{\eta_\cP}(\bI\{W_i=1\},\bI\{W_j=1\}) = & \ \bE_{\eta_\cP}\big[\bI\{W_i=1\}\bI\{W_j=1\}\big] - \bE_{\eta_\cP}[\bI\{W_i=1\}]\bE_{\eta_\cP}[\bI\{W_j=1\}] \\
= & \ \frac{ \dbinom{s_l-2}{ \frac{s_l}{2}-2 } }{ \dbinom{s_l}{ \frac{s_l}{2} } } - \frac{1}{4} \\
= & \ - \frac{1}{4 (s_l-1)}.
\end{align*}
See Table~\ref{tbl:VarCovMatrix6} for an illustration of a block of matrix.
Recall that there are $k$ strata in the partition.
We can then write the expression in \eqref{eqn:QPVarCov} as
\begin{align*}
\bm{g}^\top \bb{v} \bm{g} = \sum_{l=1}^k \Bigg\{ \frac{1}{4} \ \sum_{i \in S_l} \ \sum_{\substack{j \in S_l \\ i \ne j}} \ \frac{(g_i-g_j)^2}{s_l-1} \Bigg\}.
\end{align*}
For each $S_l$, if we re-arrange the baselines such that $g_{j_1} \geq g_{j_2} \geq ... \geq g_{j_{s_l}}$, then we can repeatedly use \eqref{eqn:MatchedPairUsefulIneq} and lower bound the double summation term by matched pairs, that is,
\begin{align*}
\sum_{i \in S_l} \ \sum_{\substack{j \in S_l \\ i \ne j}} \ \frac{(g_i-g_j)^2}{s_l-1} \geq (g_{j_1} - g_{j_2})^2 + ... + (g_{j_{s_{k-1}}} - g_{j_{s_l}})^2.
\end{align*}
This essentially breaks $\bm{g}^\top \bb{v} \bm{g}$ into size-two strata.
To finish the proof, we use \eqref{eqn:MatchedPairUsefulIneq} again to show that the optimal matched pair partition is 
\begin{align*}
\cP = \Big\{ \{1,2\}, \{3,4\}, ..., \{n-1, n\} \Big\}.
\end{align*}
\hfill \halmos
\endproof

\section{Proofs from Section~\ref{sec:Uncertainty3}}

\proof{Proof of Theorem~\ref{thm:SynthControlExp}.}
For any period $t \in \{T_0+1, ..., T\}$, we decompose $\widehat{\tau}^{SC}_t(\bm{u}^*, \bm{v}^*)$ as follows,
\begin{align*}
\widehat{\tau}^{SC}_t(\bm{u}^*, \bm{v}^*) - \tau_t = \Big( \sum_{j=1}^n u_j^* Y_{jt}(1) - \sum_{j=1}^n f_j Y_{jt}(1) \Big) - \Big( \sum_{j=1}^n v_j^* Y_{jt}(0) - \sum_{j=1}^n f_j Y_{jt}(0) \Big).
\end{align*}
We examine the above two parentheses separately.
Under the linear factor model \eqref{eqn:LinearFactorModel} we have that during the experimental periods $t \in \{T_0+1,...,T\}$,
\begin{multline}
\sum_{j=1}^n u_j^* Y_{jt}(1) - \sum_{j=1}^n f_j Y_{jt}(1) = \bm{\beta}_t(1)^\top \Big( \sum_{j=1}^n u_j^* \bm{X}_j - \sum_{j=1}^n f_j \bm{X}_j \Big) \\
+ \bm{\lambda}_t(1)^\top \Big( \sum_{j=1}^n u_j^* \bm{\mu}_j - \sum_{j=1}^n f_j \bm{\mu}_j \Big) + \Big( \sum_{j=1}^n u_j^* \epsilon_{jt}(1) - \sum_{j=1}^n f_j \epsilon_{jt}(1) \Big). \label{eqn:SC1}
\end{multline}
In the above expression \eqref{eqn:SC1}, we can fit the observed covariates so the first term is small.
The third term consists of mean-zero random noises.
The second term consists of unobserved covariates, which we never directly observe.
But under the linear factor model \eqref{eqn:LinearFactorModel} we can fit the unobserved covariates by fitting the historical observed outcomes.

Denote $\bm{Y}_j(0)$ to be a $T_0$-dimensional column vector whose $t$-th element is equal to $Y_{jt}(0)$.
Recall that $\bblambda(0)$ is a $(T_0 \times r)$ matrix whose $t$-th row is equal to $\bm{\lambda}_t(0)^\top$.
Denote $\bbbeta(0)$ to be a $(T_0 \times d)$ matrix whose $t$-th row is equal to $\bm{\beta}_t(0)^\top$.
Denote $\bm{\epsilon}_j(0)$ to be a $T_0$-dimensional column vector whose $t$-th element is equal to $\epsilon_{jt}(0)$.
Under the linear factor model \eqref{eqn:LinearFactorModel} we have that during the pre-experimental periods $t\in[T_0]$,
\begin{multline*}
\sum_{j=1}^n u_j^* \bm{Y}_j(0) - \sum_{j=1}^n f_j \bm{Y}_j(0) = \bbbeta(0) \Big( \sum_{j=1}^n u_j^* \bm{X}_j - \sum_{j=1}^n f_j \bm{X}_j \Big) \\
+ \bblambda(0) \Big( \sum_{j=1}^n u_j^* \bm{\mu}_j - \sum_{j=1}^n f_j \bm{\mu}_j \Big) + \Big( \sum_{j=1}^n u_j^* \bm{\epsilon}_j(0) - \sum_{j=1}^n f_j \bm{\epsilon}_j(0) \Big).
\end{multline*}
Pre-multiplying $\bm{\lambda}_t(1)^\top \big( \bblambda(0)^\top \bblambda(0) \big)^{-1} \bblambda(0)^\top$ yields
\begin{align}
\bm{\lambda}_t(1)^\top \big( \bblambda(0)^\top \bblambda(0) \big)^{-1} \bblambda(0)^\top & \Big(\sum_{j=1}^n u_j^* \bm{Y}_j(0) - \sum_{j=1}^n f_j \bm{Y}_j(0)\Big) \label{eqn:SC2} \\
= & \ \bm{\lambda}_t(1)^\top \big( \bblambda(0)^\top \bblambda(0) \big)^{-1} \bblambda(0)^\top \bbbeta(0) \Big( \sum_{j=1}^n u_j^* \bm{X}_j - \sum_{j=1}^n f_j \bm{X}_j \Big) \nonumber \\
& + \bm{\lambda}_t(1)^\top \Big( \sum_{j=1}^n u_j^* \bm{\mu}_j - \sum_{j=1}^n f_j \bm{\mu}_j \Big) \nonumber \\
& + \bm{\lambda}_t(1)^\top \big( \bblambda(0)^\top \bblambda(0) \big)^{-1} \bblambda(0)^\top \Big( \sum_{j=1}^n u_j^* \bm{\epsilon}_j(0) - \sum_{j=1}^n f_j \bm{\epsilon}_j(0) \Big). \nonumber
\end{align}

Combining \eqref{eqn:SC1} and \eqref{eqn:SC2} we have
\begin{align}
\sum_{j=1}^n u_j^* Y_{jt}(1) - \sum_{j=1}^n f_j Y_{jt}(1) = & \ \Big(\bm{\beta}_t(1)^\top - \bm{\lambda}_t(1)^\top \big( \bblambda(0)^\top \bblambda(0) \big)^{-1} \bblambda(0)^\top \bbbeta(0)\Big) \Big( \sum_{j=1}^n u_j^* \bm{X}_j - \sum_{j=1}^n f_j \bm{X}_j \Big) \label{eqn:SC3} \\
& + \bm{\lambda}_t(1)^\top \big( \bblambda(0)^\top \bblambda(0) \big)^{-1} \bblambda(0)^\top \Big(\sum_{j=1}^n u_j^* \bm{Y}_j(0) - \sum_{j=1}^n f_j \bm{Y}_j(0)\Big) \nonumber \\
& - \bm{\lambda}_t(1)^\top \big( \bblambda(0)^\top \bblambda(0) \big)^{-1} \bblambda(0)^\top \sum_{j=1}^n u_j^* \bm{\epsilon}_j(0) \nonumber \\
& + \bm{\lambda}_t(1)^\top \big( \bblambda(0)^\top \bblambda(0) \big)^{-1} \bblambda(0)^\top \sum_{j=1}^n f_j \bm{\epsilon}_j(0) \nonumber \\
& + \Big( \sum_{j=1}^n u_j^* \epsilon_{jt}(1) - \sum_{j=1}^n f_j \epsilon_{jt}(1) \Big). \nonumber
\end{align}

Below we bound each line in \eqref{eqn:SC3}.
Using Cauchy–Schwarz inequality and the eigenvalue bound on the Rayleigh quotient, for any $t, s \in [T]$ and $w \in \{0,1\}$,
\begin{multline*}
\left| \bm{\lambda}_t(w)^\top \big(\bblambda(0)^\top \bblambda(0)\big)^{-1} \bm{\lambda}_s(0) \right| \\
\leq \Big( \bm{\lambda}_t(w)^\top \big(\bblambda(0)^\top \bblambda(0)\big)^{-1} \bm{\lambda}_t(w) \Big)^{\frac{1}{2}} \Big( \bm{\lambda}_s(0)^\top \big(\bblambda(0)^\top \bblambda(0)\big)^{-1} \bm{\lambda}_s(0) \Big)^{\frac{1}{2}} \leq \frac{ \overline{\lambda}^2 r}{T_0 \underline{\zeta}}.
\end{multline*}
Therefore, the absolute value of each element in vector $\big(\bm{\beta}_t(1)^\top - \bm{\lambda}_t(1)^\top \big(\bblambda(0)^\top \bblambda(0)\big)^{-1} \bblambda(0) \bbbeta(0)\big)$ is bounded by $\overline{\beta}\big(1 + \dfrac{\overline{\lambda}^2 r}{\underline{\zeta}}\big)$.
We use Cauchy–Schwarz inequality and upper bound the first line of \eqref{eqn:SC3}
\begin{align*}
\bigg\vert\Big(\bm{\beta}_t(1)^\top - \bm{\lambda}_t(1)^\top \big( \bblambda(0)^\top \bblambda(0) \big)^{-1} \bblambda(0)^\top \bbbeta(0)\Big) & \Big( \sum_{j=1}^n u_j^* \bm{X}_j - \sum_{j=1}^n f_j \bm{X}_j \Big) \bigg\vert\\
\leq & \ \overline{\beta}\big(1 + \frac{\overline{\lambda}^2 r}{\underline{\zeta}}\big) d^\frac{1}{2} \cdot \bigg\|\sum_{j=1}^n u^*_j \bm{X}_j - \sum_{j=1}^n f_j \bm{X}_j\bigg\|_2 \\
\leq & \ \overline{\beta}\big(1 + \frac{\overline{\lambda}^2 r}{\underline{\zeta}}\big) d c,
\end{align*}
and similarly the second line of \eqref{eqn:SC3}
\begin{align*}
\bigg\vert \bm{\lambda}_t(1)^\top \big( \bblambda(0)^\top \bblambda(0) \big)^{-1} \bblambda(0)^\top \Big(\sum_{j=1}^n u_j^* \bm{Y}_j(0) - \sum_{j=1}^n f_j \bm{Y}_j(0)\Big) \bigg\vert \leq \frac{\overline{\lambda}^2 r}{T_0 \underline{\zeta}} T_0^{\frac{1}{2}} \cdot T_0^{\frac{1}{2}} c \leq \frac{\overline{\lambda}^2 r}{\underline{\zeta}} c.
\end{align*}
Then, for the third line of \eqref{eqn:SC3}, we denote
\begin{align*}
\xi_{jt} = \bm{\lambda}_t(1)^\top \big( \bblambda(0)^\top \bblambda(0) \big)^{-1} \bblambda(0)^\top \bm{\epsilon}_j(0) = \sum_{s=1}^{T_0} \bm{\lambda}_t(1)^\top \big( \bblambda(0)^\top \bblambda(0) \big)^{-1} \bblambda(0)^\top \epsilon_{js}(0).
\end{align*}
Because $\xi_{jt}$ is a linear combination of independent sub-Gaussian random variables with variance proxy $\overline{\sigma}^2$, $\xi_{jt}$ must be a sub-Gaussian random variable with variance proxy $\big(\frac{\overline{\lambda}^2 r}{T_0 \underline{\zeta}}\big)^2 \overline{\sigma}^2 T_0 = \big(\frac{\overline{\lambda}^2 r}{\underline{\zeta}}\big)^2 \frac{\overline{\sigma}^2}{T_0}$.
Let $\mathcal{S} = \{\bm{u} \vert \sum_{j=1}^n u_j = 1 \}$ be the unit simplex.
Using Theorem~1.16 from \citet{rigollet2019high} we have
\begin{align*}
\bigg\vert \bE\Big[-\sum_{j=1}^n u_j^* \xi_{jt}\Big] \bigg\vert \leq \bE\bigg[ \max_{\bm{u} \in \mathcal{S}} \Big\vert \sum_{j=1}^n u_j \xi_{jt} \Big\vert \bigg] \leq \frac{\overline{\lambda}^2 r}{\underline{\zeta}} (2 \log{2n})^\frac{1}{2} \ \overline{\sigma} \ T_0^{-\frac{1}{2}}.
\end{align*}
Finally, the last two lines of \eqref{eqn:SC3} are mean zero.
Putting them all back to \eqref{eqn:SC3} we have
\begin{align*}
\bigg\vert \bE\Big[\sum_{j=1}^n u_j^* Y_{jt}(1) - \sum_{j=1}^n f_j Y_{jt}(1)\Big] \bigg\vert \leq \bigg( \overline{\beta} d + \big(1 + \overline{\beta} d \big) \frac{\overline{\lambda}^2 r}{\underline{\zeta}} \bigg) c + \frac{\overline{\lambda}^2 r}{\underline{\zeta}} (2 \log{2n})^\frac{1}{2} \ \overline{\sigma} \ T_0^{-\frac{1}{2}}.
\end{align*}
Similarly for the synthetic control unit, we have
\begin{align*}
\bigg\vert \bE\Big[\sum_{j=1}^n v_j^* Y_{jt}(0) - \sum_{j=1}^n f_j Y_{jt}(0)\Big] \bigg\vert \leq \bigg( \overline{\beta} d + \big(1 + \overline{\beta} d \big) \frac{\overline{\lambda}^2 r}{\underline{\zeta}} \bigg) c + \frac{\overline{\lambda}^2 r}{\underline{\zeta}} (2 \log{2n})^\frac{1}{2} \ \overline{\sigma} \ T_0^{-\frac{1}{2}}.
\end{align*}
Combining both parts we finish the proof.
\hfill \halmos
\endproof

\end{APPENDIX}

%%%%%%%%%%%%%%%%
\end{document}